\documentclass{appolb}
\usepackage{epsf,amsmath}
\usepackage{graphics}

\begin{document}

\title{Structure and evolution of the foreign exchange networks}

\author{Jaros{\l}aw Kwapie\'n$^1$, Sylwia Gworek$^1$, Stanis{\l}aw 
Dro\.zd\.z$^{1,2}$
\address{$^1$ Institute of Nuclear Physics, Polish Academy of Sciences,
PL--31-342 Krak\'ow, Poland \\ 
$^2$ Institute of Physics, University of Rzesz\'ow, PL--35-959 Rzesz\'ow, 
Poland}}

\maketitle

\date{Received: date / Revised version: date}

\abstract{
We investigate topology and temporal evolution of the foreign currency 
exchange market viewed from a weighted network perspective. Based on 
exchange rates for a set of 46 currencies (including precious metals), we 
construct different representations of the FX network depending on a 
choice of the base currency. Our results show that the network structure 
is not stable in time, but there are main clusters of currencies, which 
persist for a long period of time despite the fact that their size and 
content are variable. We find a long-term trend in the network's evolution 
which affects the USD and EUR nodes. In all the network representations, 
the USD node gradually loses its centrality, while, on contrary, the EUR 
node has become slightly more central than it used to be in its early 
years. Despite this directional trend, the overall evolution of the 
network is noisy.}


\section{Introduction}

A network-based approach to analysis of the foreign exchange (FX)  market 
has relatively short history in econophysics, and despite its usefulness 
in describing market properties, only a few papers were published on this 
subject. Ortega and Matesanz~\cite{ortega06} analyzed a set of exchange 
rates of 28 currencies recorded over the 1990-2002 period and by means of 
the minimal spanning tree (MST) and the ultrametric distance they 
identified a hierarchical structure of the FX network, in which the global 
market is subdivided into sectors comprising countries from the same 
geographical regions. A similar conclusion was drawn independently by 
Mizuno et al.~\cite{mizuno06} who studied data for 26 currencies and 3 
precious metals for a shorter time interval 1999-2003. Their results 
confirmed a leading role played by USD in the market and showed the 
existence of a few other key currencies dominating regionally. They are 
typically associated with the largest regional economies like euro and the 
Australian dollar. Both the above-described studies were based on networks 
with a fixed base currency.

A different approach was adopted by McDonald et al.~\cite{mcdonald05} who 
investigated a network of all possible exchange rates for 10 currencies 
and gold (years 1993-94 and 2004) without expressing the data in the same 
base currency. They found that topological properties of the 
so-constructed FX network evidently differ from the properties of a 
network built on surrogate data. They also addressed a problem of 
stability of this network and found that about a half of the links in MST 
survive for as long as two years. Moreover, despite the fact that other 
links frequently appear and disappear, there are clusters of MST nodes 
which can emerge and survive for finite periods of time but still being 
subject to some long-term evolution - a behaviour that closely resembles 
the ecological systems. A specific example of such a behaviour was given 
later by Naylor et al.~\cite{naylor07}, who analyzed temporal evolution of 
two currency networks constructed from 44 currencies in the time interval 
1995-2001 and based on the New Zealand dollar and the US dollar. By using 
both MST and the ultrametric distance hierarchical tree, the authors of 
ref.~\cite{naylor07} showed changes in the network structure during the 
Asian crisis period 1997-1998 that manifested themselves by a transient 
emergence of a cluster of strongly coupled South-East Asian currencies.

Finally, in their two consecutive papers considering a large set of 57 
currencies and 3 precious metals from the interval 1999-2005, Dro\.zd\.z 
and coworkers~\cite{drozdz07,gorski08} discussed the structure of the 
forex networks created for different choices of the base currency. They 
showed that the network structure is significantly base-dependent and can 
vary from hierarchical and scale-free for a majority of base currencies to 
almost random for the US dollar and its pegged satellites. Also the 
average coupling between pairs of exchange rates, expressed by, e.g., the 
largest eigenvalue of the corresponding correlation matrix, is 
anticorrelated with the base currency's role in the market: the largest 
eigenvalue has the highest magnitude for marginal currencies and precious 
metals, and the smallest magnitude for the US dollar.

In the present work we systematically look at the evolution of the 
currency network and analyze changes in its structure with time. Our 
data~\cite{sauder08} comprise the daily cross-rates for 43 independent 
currencies listed in footnote\footnote{AUD, BRL, CAD, CHF, CLP, COP, CZK, 
DZD, EGP, EUR, FJD, GBP, GHS, HNL, HUF, IDR, ILS, INR, ISK, JMD, JPY, KRW, 
LKR, MAD, MXN, NOK, NZD, PEN, PHP, PKR, PLN, RON, RUB, SDD, SEK, SGD, THB, 
TND, TRY, TWD, USD, ZAR, ZMK.} and 3 precious metals (XAU, XAG, XPT). By 
``independent currencies'' we mean currencies which are not explicitely 
pegged to any other currency. According to this condition, we had to 
exclude from our analysis a few liquid currencies like, e.g., the 
Malaysian ringgit, the Hong Kong dollar (both pegged to the US dollar), 
and the Danish krone (pegged to euro via ERM II). The data under study 
spans a time interval of 9.5 years from 12/15/1998 (when euro was 
introduced) to 06/30/2008.

\section{Methods}

For each exchange rate B/X expressing a unit of a base currency B in units
of another currency X, we consider a time series of normalized (to unit 
variance and zero mean) logarithmic returns $g^{\rm B}_{\rm X}(i)$, where 
$i = 1,...,T$ denotes consecutive trading days ($T=2394$). In this way for 
a complete set of 46 currencies we obtain 2070 different synchronous time 
series. All the signals were preprocessed in order to eliminate artifacts.
Also a filter was applied to eliminate extreme returns which can potentially
dominate the outcomes (in a consequence, all the returns exceeding 
$\pm 10\sigma$ were replaced by one of these threshold values).

In principle, it would be possible to study a complete network from the
full set of signals (as it was done in~\cite{mcdonald05}) but such an
approach: (i)  would lead to results which are difficult to interpret, and
(ii) we would not have an opportunity to change a reference frame and
analyze the network from a point of view of different base currencies.
Therefore we follow an alternative
approach~\cite{ortega06,mizuno06,naylor07,drozdz07,gorski08} in which we
construct a currency network by analyzing only those signals, which share
the same base currency B. By changing B, we obtain different
representations of this network, which opens space for a more subtle 
analysis of the market structure.

Technical details of the construction of a network are as follows. Let us 
denote by $N$ the number of time series with the same base currency 
($N=45$, independently of B). For a given B we create an $N \times T$ data 
matrix ${\bf M}^{\rm B}$ and then an $N \times N$ correlation matrix 
${\bf R}^{\rm B}$ according to the formula:
\begin{equation}
{\bf R}^{\rm B} = {1 \over T} {\bf M}^{\rm B} \tilde{\bf M}^{\rm B} ,
\end{equation}
where $\tilde{\cdot}$ stands for matrix transpose. Entries $R^{\rm B}_{\rm 
X,Y}$ of the correlation matrix are the Pearson correlation coefficients 
quantifying linear dependencies between the pairs of time series 
associated with the B/X and B/Y exchange rates.  ${\bf R}^{\rm B}$ 
completely defines the structure of the B-based currency network which is 
a fully connected, undirected, weighted network with $N$ nodes 
representing the exchange rates B/X, and $N(N-1)/2$ internode connections 
with weights $\omega^{\rm B}_{\rm X,Y} = |R^{\rm B}_{\rm X,Y}|$.

Such a complete network can in general be presented on a graph, but 
showing all the connections would lead to an unreadable picture even for 
small networks. A more appropriate method in this context is application 
of a minimal spanning tree graph~\cite{mantegna99}. To obtain MST, we 
derive a distance matrix ${\bf D}^{\rm B}$ with entries $d^{\rm B}_{\rm 
X,Y}$ defined by
\begin{equation}
d^{\rm B}_{\rm X,Y} = \sqrt{2(1 - R^{\rm B}_{\rm X,Y})}, \ \ \ i = 1,...,N 
- 1.
\label{distance}
\end{equation}
For a pair of identical signals $d^{\rm B}_{\rm X,Y} = 0$, while for a
pair of statistically uncorrelated ones $d^{\rm B}_{\rm X,Y} = \sqrt{2}$.
Anticorrelations are expressed by $\sqrt{2} < d^{\rm B}_{\rm X,Y} \le 2$. 
MST is now constructed by sorting the entries of ${\bf D}^{\rm B}$ and, in 
each step of the construction, by connecting the closest nodes with 
respect to $d^{\rm B}_{\rm X,Y}$ in such a manner that no node is 
connected via more than one path and no node is left alone. More detailed 
instructions can be found, e.g., in~\cite{mantegna99}. A complete MST 
graph consists of $N$ nodes and $N - 1$ edges, each edge connecting a node 
with its closest neighbour.

In this way we can reduce the number of connections and make a graphical 
representation of the network to be much more readable. However, while 
going beyond the graphical convenience of using trees instead of complete 
networks, and considering also the topological characteristics of MSTs, 
one faces a problem of adequacy of this network representation in the case 
of financial data. This problem stems from the fact that an MST does not 
have an obvious economical interpretation, which could naturally favour it 
over other possible graphs. Nevertheless, as we show later, the results of our
analysis of the MST's topology go in parallel with those based on an analysis
of the complete network. This observation together with the simplicity of the 
MST construction both justify the possibility of restricting a study of the 
forex network's topology to minimal spanning trees and extend validity of the
so-obtained conclusions over the whole network.

\section{Results and discussion}

Before we analyze temporal evolution of the FX network, let us start with 
a description of its average structure over the full period 12/15/1998 - 
06/30/2008. First, we need to choose a representation of the network, i.e. 
to choose B. Selecting B means attaching to this currency a reference 
frame and, effectively, exclude it from the network. In order to obtain a 
network with a topology as close to the real market as possible, we have 
to single out a currency which is of marginal importance, i.e. a currency 
whose exchange rates to other currencies have minimal influence on other 
exchange rates not involving this currency. Good candidates are precious 
metals and some third world exotic currencies with idiosyncratic dynamics. 
In terms of the eigenspectrum of ${\bf R}^{\rm B}$, such currencies 
develop an extremely large ``energy gap'' (see~\cite{drozdz07}).

\begin{figure}
\hspace{2.0cm}
\epsfxsize 6cm
\epsffile{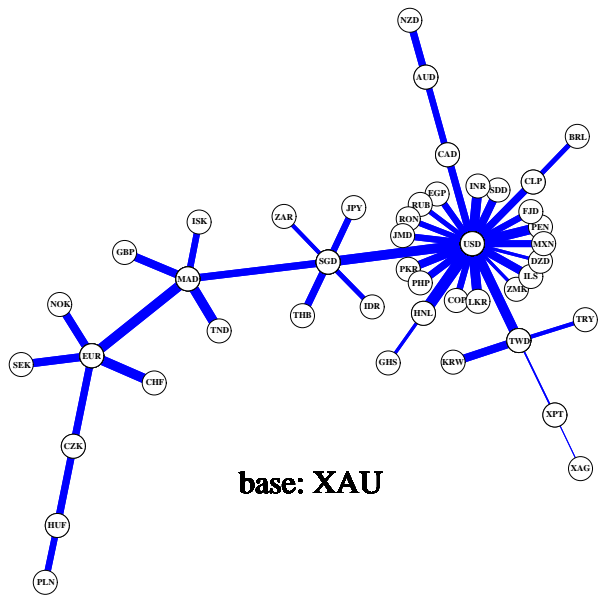}
\caption{Minimal Spanning Tree with weighted edges for the XAU-based FX   
network representation calculated for the full interval
12/15/1998-06/30/2008. Line widths are proportional to weights $\omega$ of 
the corresponding edges, while line lengths are arbitrary.}
\end{figure}

Figure 1 shows the weighted MST for gold (B $\equiv$ XAU). (Due to the 
fact that in this and in other MST graphs all the cross-rates share the 
same base currency, for clarity we denote the nodes only by their price 
currencies.) Clearly, the network in Figure 1 is dominated by the USD node 
with the highest degree $K = 21$. Other distinguished nodes are SGD ($K = 
6$), EUR, MAD ($K = 5$), and TWD ($K = 4$), but none of them can be 
compared with USD. This supremacy of the USD node over the network is not 
surprising: according to the Bank of International Settlements 2007 
study~\cite{bis07}, transactions involving USD account for 86.3\% of the 
global FX turnover, while the same index is 37.0\% for euro, 16.5\% for 
yen, and 15.0\% for the British pound (both exchange directions were taken 
into account, thus the numbers for all currencies sum up to 200\%). As one 
can see, despite the above-discussed fact that from a purely economical 
point of view an MST might at first sight seem to be a rather abstract 
object, actually its topology can reflect some important properties of the 
real FX market.

An even more credible topological measure of node centrality in the 
B-based network is the betweenness $b^{\rm B}({\rm X})$ of a node X, which 
for an unweighted (binary) MST graph is given by a simple formula:
\begin{equation}
b^{\rm B}({\rm X}) = \sum_{{\rm Y,Z}: {\rm Y} \neq {\rm Z}} {\delta^{\rm
B}_{\rm Y,Z}({\rm X}) \over (N - 1)(N - 2)},
\end{equation}
where $\delta^{\rm B}_{\rm Y,Z}({\rm X})$ equals 1 if the path linking a
pair of nodes (Y,Z) goes through X, or equals 0 otherwise. The betweenness
thus quantifies a fraction of node pairs that are connected via a node X
with respect to the total number of possible pairs (Y,Z) which is equal to
$(N-1)(N-2)$. This version of betweenness can be generalized for a
weighted MST, but in our case the above-defined topological version is
sufficient to characterize the most important properties of the trees.
The so-defined betweenness of the main nodes for the XAU-based network has
the following values: 0.82 (USD), 0.49 (SGD), 0.37 (MAD), 0.24 (EUR), and
0.17 (TWD). It should be noted, however, that this quantity (together with
the node degree $K$) characterizes the structure of the network, while
does not necessarily reflect the importance of particular nodes in the
global financial system. For example, the XAU/SGD and XAU/MAD cross-rates
are rather exotic and their significant role in the MST in Figure 1 stems
mainly from their intermediate location between USD and EUR $-$ the actual
hubs of the FX market.

\begin{figure}[h]
\hspace{2.0cm}
\epsfxsize 6cm
\epsffile{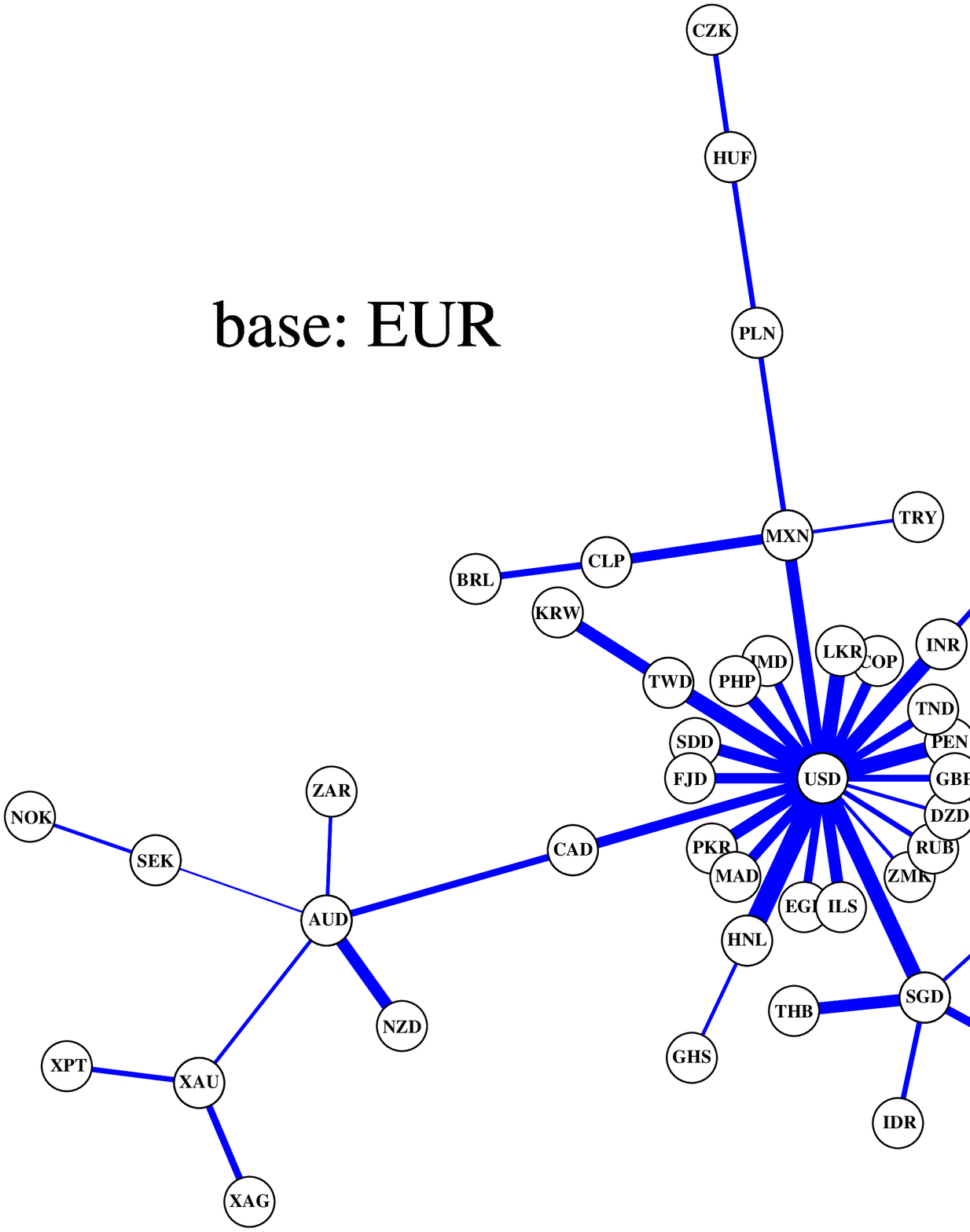}

\vspace{0.5cm}
\hspace{2.0cm}
\epsfxsize 6cm
\epsffile{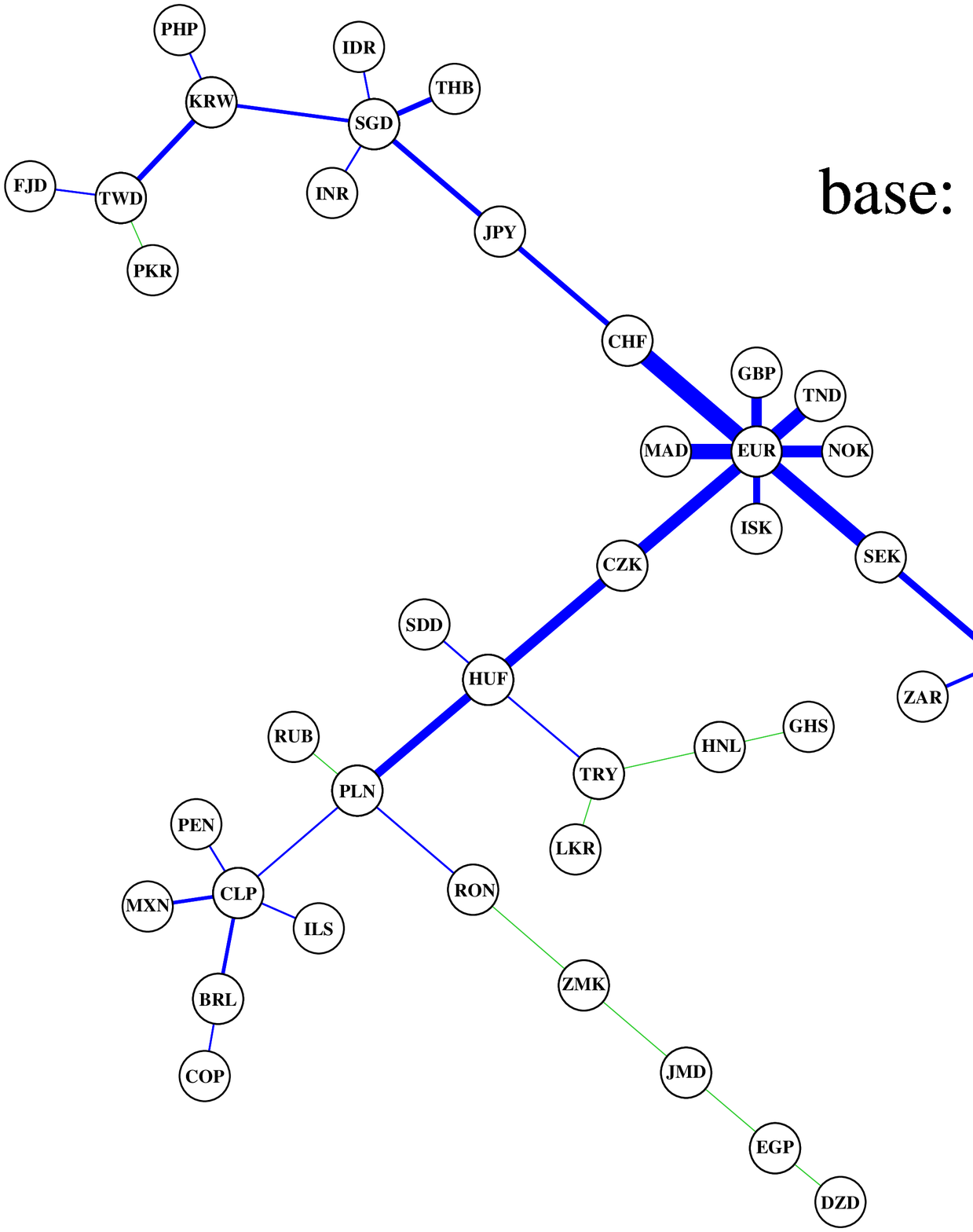}
\caption{Minimal Spanning Trees for the EUR-based network (top panel)
and for the USD-based network (bottom panel) calculated for the full 
interval 12/15/1998-06/30/2008. Green edges connect anticorrelated 
nodes.}
\end{figure}

The MST edges in Figure 1 have widths proportional to their weights. Since
almost all the edges are thick, it is straightforward to conclude that the
XAU-based network is especially strongly coupled forming a single
``supercluster''. However, a finer structure of the XAU-based network, in
an agreement with the hierarchical nature~\cite{ravasz03} of the currency
networks, can also be seen: there exists the USD-centered cluster
combining the Latin American and Asian currencies, and the EUR-led
European cluster joined by MAD and TND. Only the nodes corresponding to
XAG and XPT are weakly attached to the rest of MST. This agrees with the
obvious observation that the prices of XAU and other precious metals have
their own dynamics different from the proper FX market. The narrow edge
linking silver and platinum in Figure 1 reflects thus only their residual
(secondary) coupling after the primary couplings inside the precious
metals group were removed by selecting XAU as the base currency. The
geographically driven cluster structure of MST supports the outcomes of
the already cited ref.~\cite{ortega06,mizuno06,naylor07}.

In a more formal way, clustering properties of a weighted network can be
characterized by the average weighted clustering coefficient, describing
the average triangle structure of edges~\cite{onnela05}:
\begin{equation}
\tilde{C}^{\rm B} = (1/N) \sum_{\rm X} \tilde{c}^{\rm B}({\rm X}),
\end{equation}
where $\tilde{c}^{\rm B}({\rm X})$ is defined by
\begin{eqnarray} 
\tilde{c}^{\rm B}({\rm X}) = {1 \over K^{\rm B}_{\rm X} (K^{\rm B}_{\rm X} 
- 1)} \sum_{\rm Y,Z} (\tilde{\omega}^{\rm B}_{\rm X,Y} \tilde{\omega}^{\rm 
B}_{\rm Y,Z} \tilde{\omega}^{\rm B}_{\rm Z,X})^{1/3}, \nonumber\\ 
\tilde{\omega}^{\rm B}_{\rm P,Q} = {\omega^{\rm B}_{\rm P,Q} \over 
\underset{\rm P,Q}{\max} (\omega^{\rm B}_{\rm P,Q})}.
\end{eqnarray}
Here $K^{\rm B}_{\rm X}$ is the degree of a node X. Obviously, 
$\tilde{c}^{\rm B}({\rm X})$ has to be calculated for a complete network, 
not for its MST representation (which does not comprise any triangles).
For the XAU-based network under study the average weighted clustering 
coefficient is high: $\tilde{C}^{\rm XAU} = 0.71$.

\begin{figure}[t]
\hspace{0.0cm}
\epsfxsize 5.5cm
\epsffile{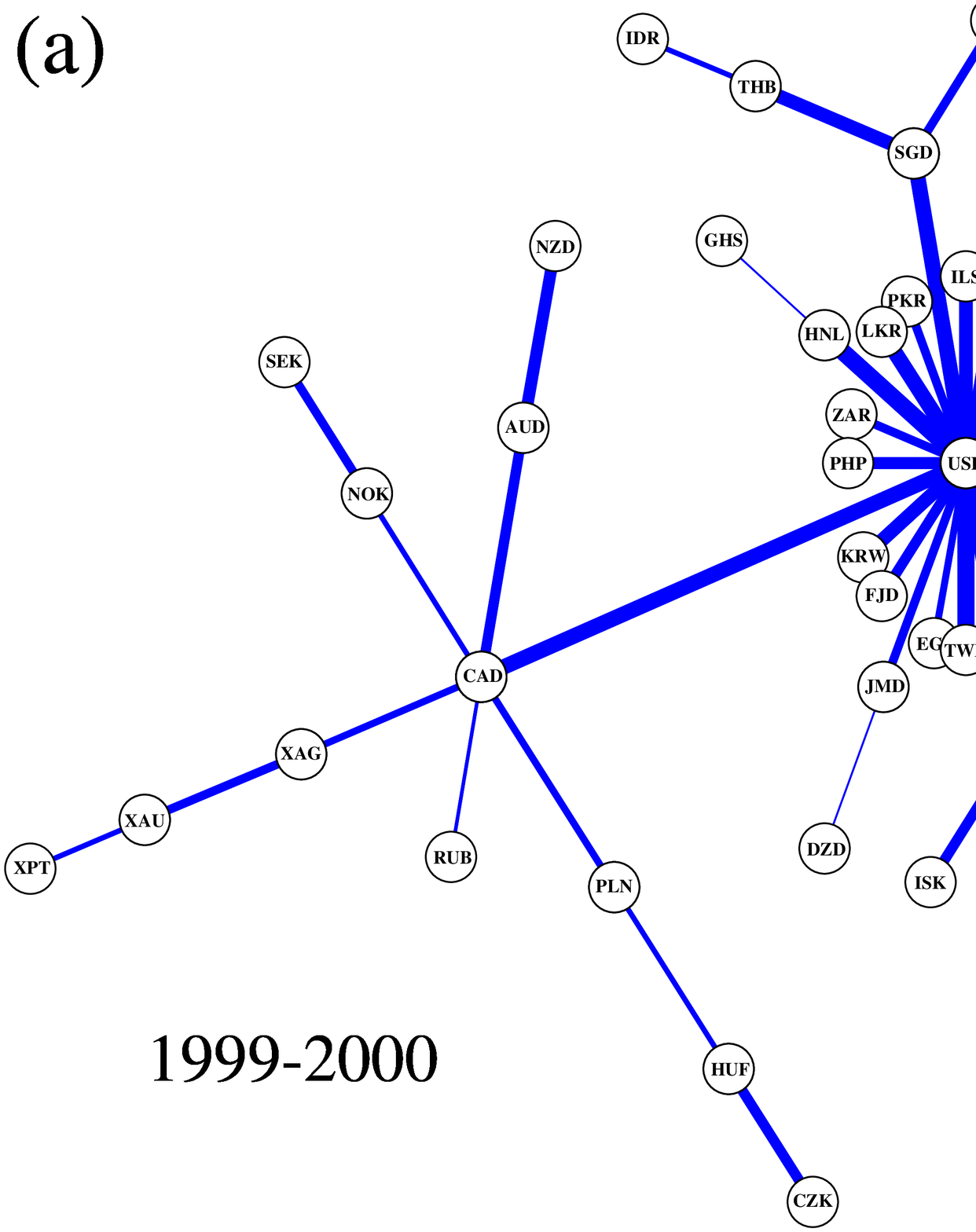}

\vspace{-1.0cm}
\hspace{6.0cm}
\epsfxsize 5.5cm
\epsffile{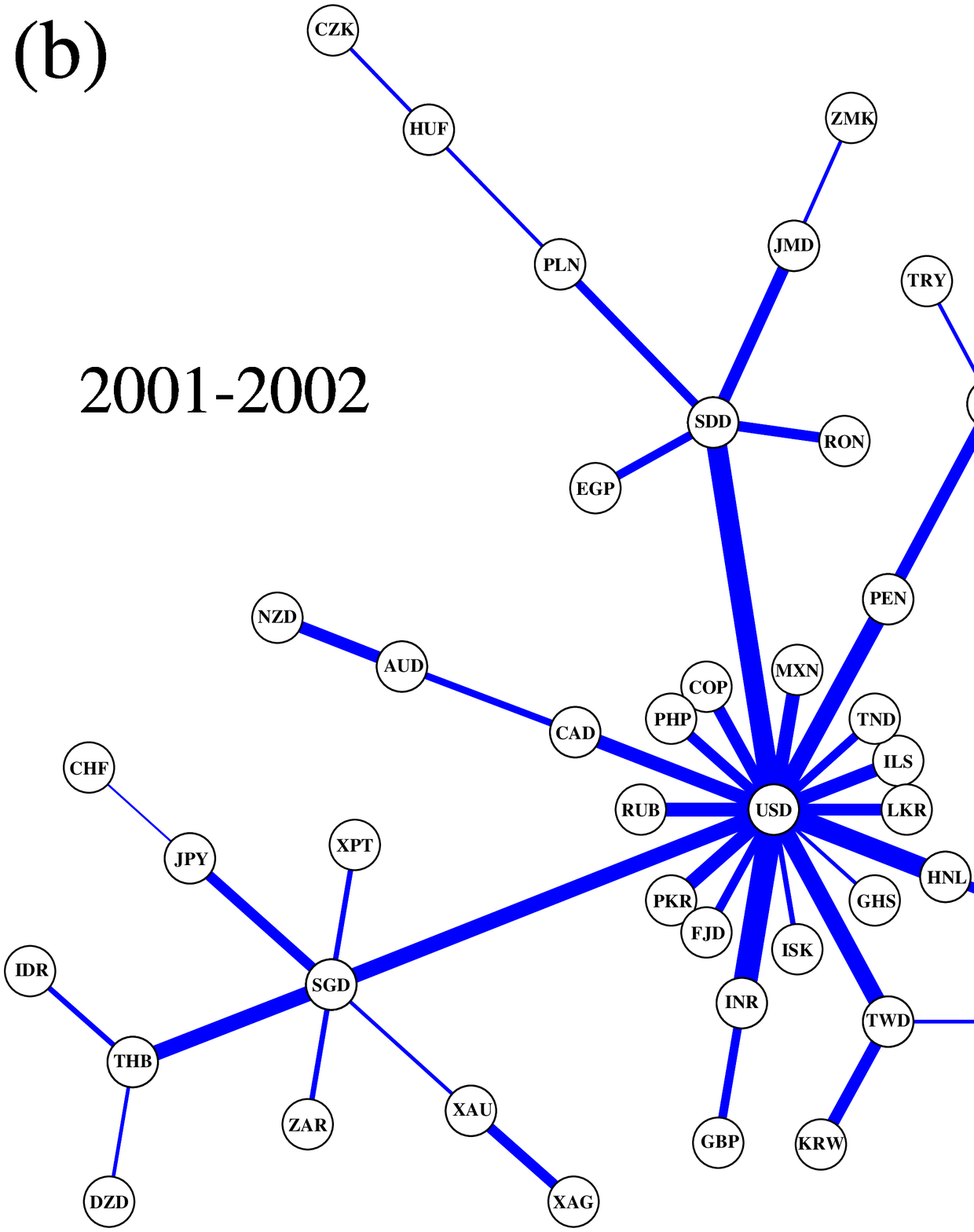}

\vspace{0.5cm}
\hspace{0.0cm}
\epsfxsize 5.5cm
\epsffile{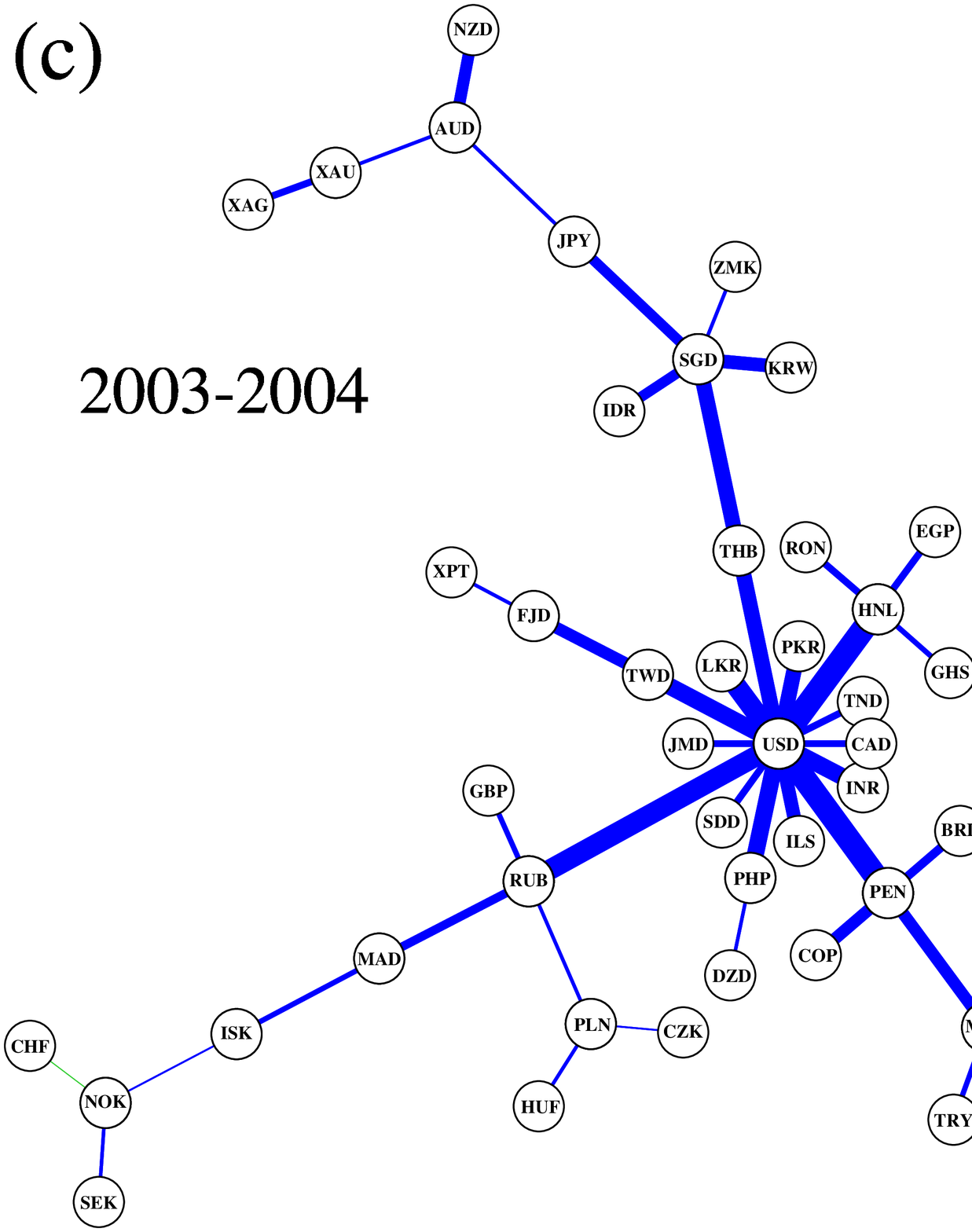}

\end{figure}

\begin{figure}
\hspace{2.0cm}
\epsfxsize 5.5cm
\epsffile{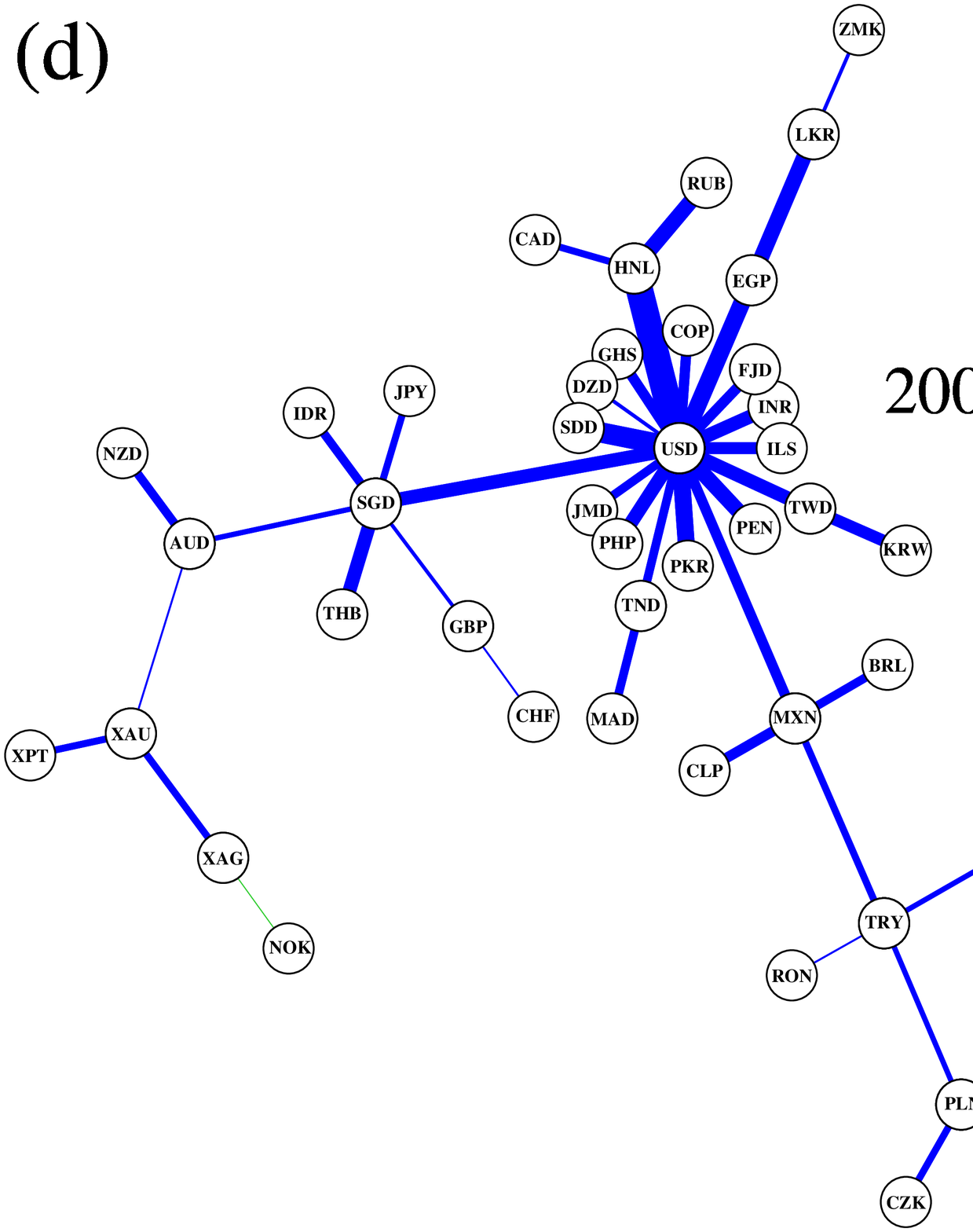}

\vspace{0.5cm}
\hspace{2.0cm}
\epsfxsize 5.5cm
\epsffile{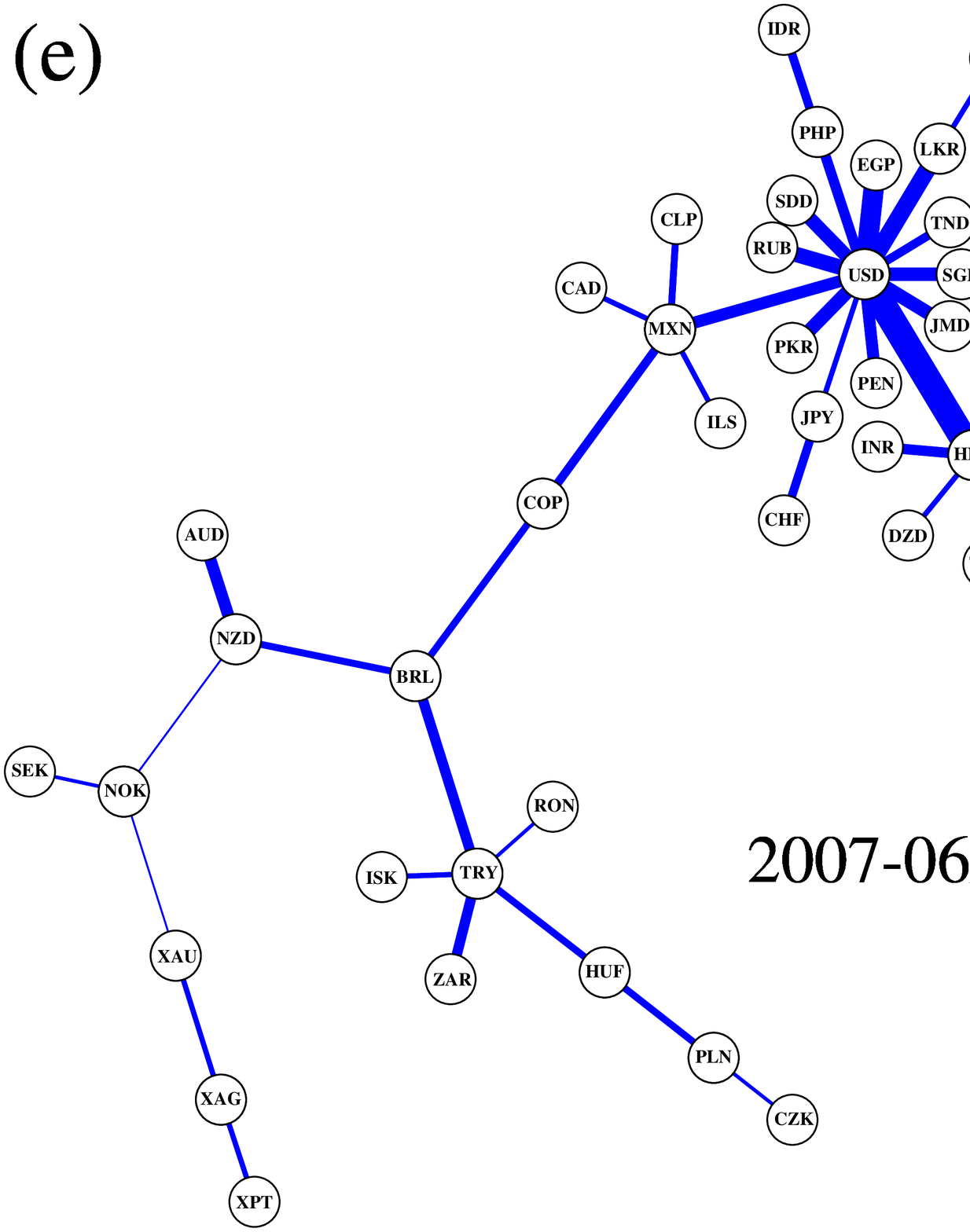}
\caption{Minimal Spanning Trees for the EUR-based network calculated in 
subintervals of 2 years (a)-(d) and 1.5 year (e). Green edges connect 
anticorrelated nodes.}
\end{figure}

\begin{figure}
\hspace{0.0cm}
\epsfxsize 5.5cm
\epsffile{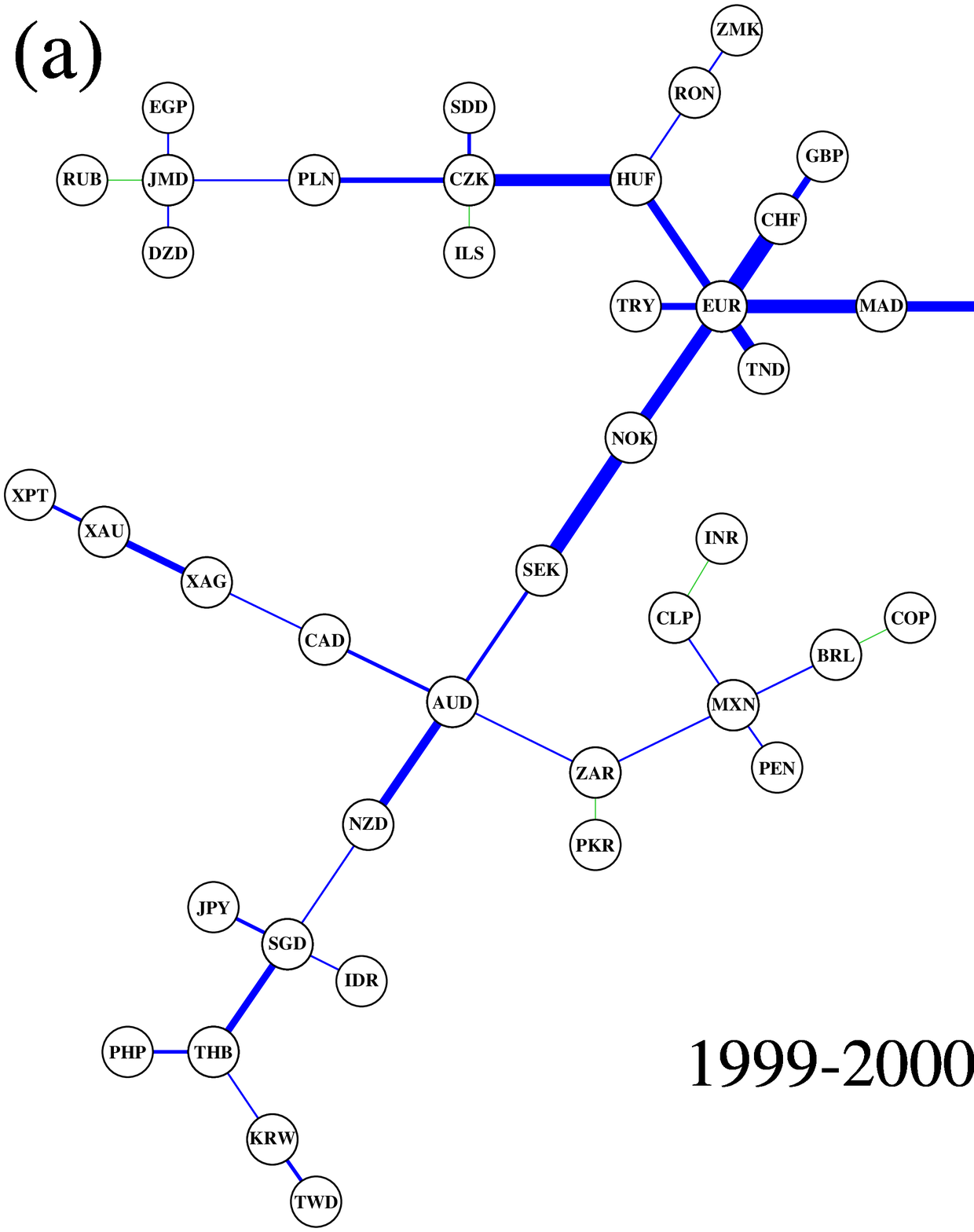}

\vspace{0.0cm}
\hspace{6.0cm}
\epsfxsize 5.5cm
\epsffile{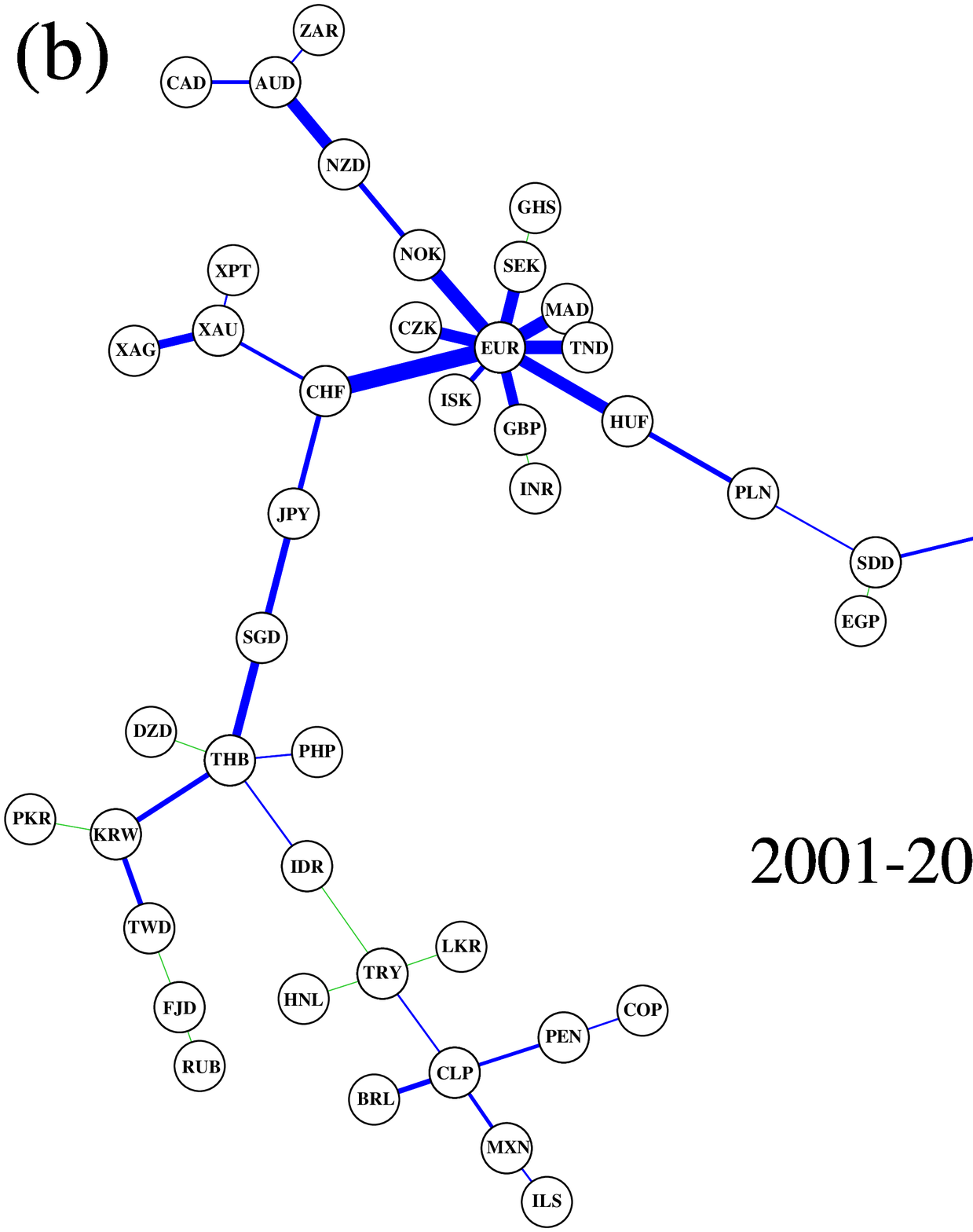}

\vspace{0.0cm}
\hspace{0.0cm}
\epsfxsize 5.5cm
\epsffile{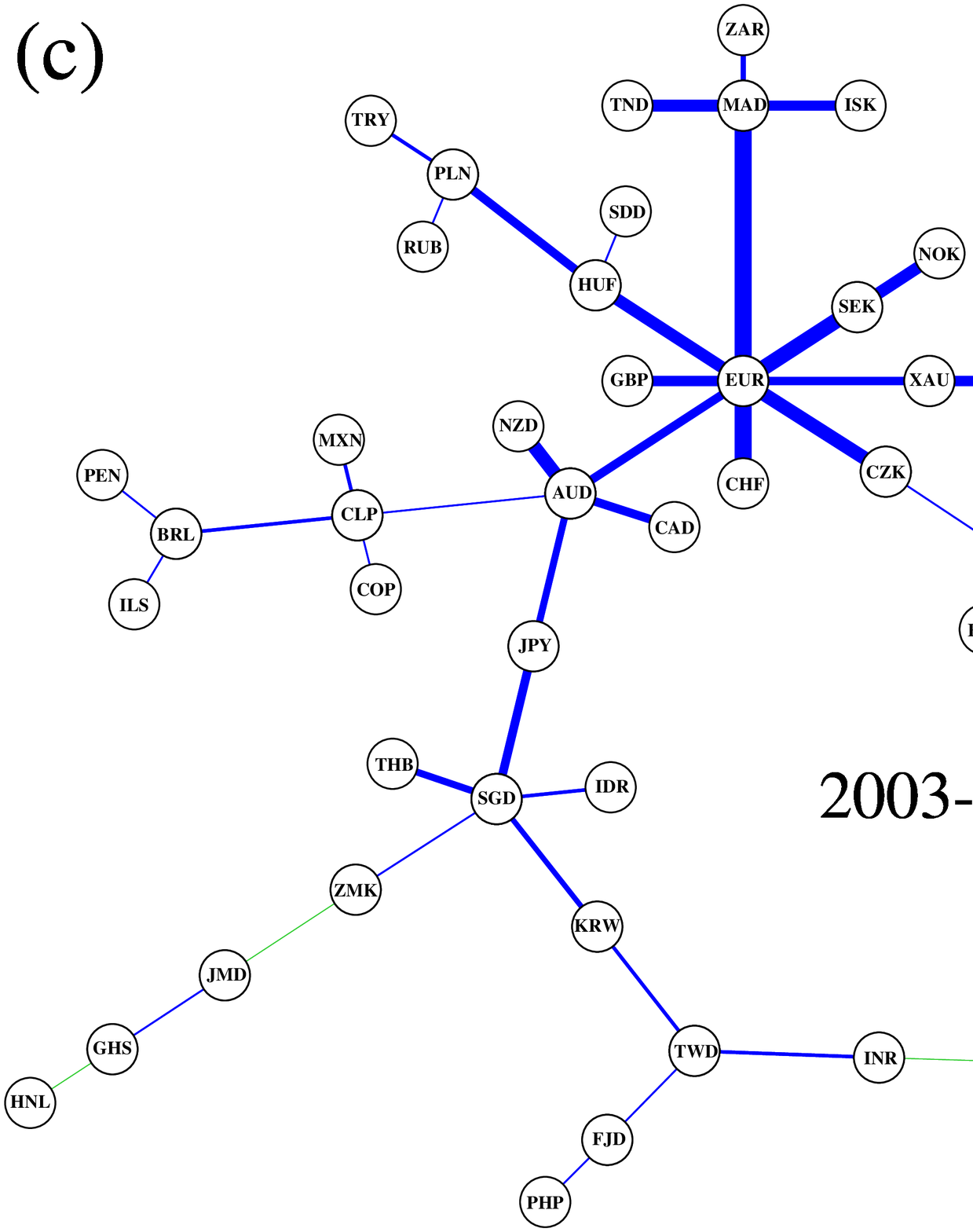}

\end{figure}

\begin{figure}
\hspace{2.0cm}
\epsfxsize 5.5cm
\epsffile{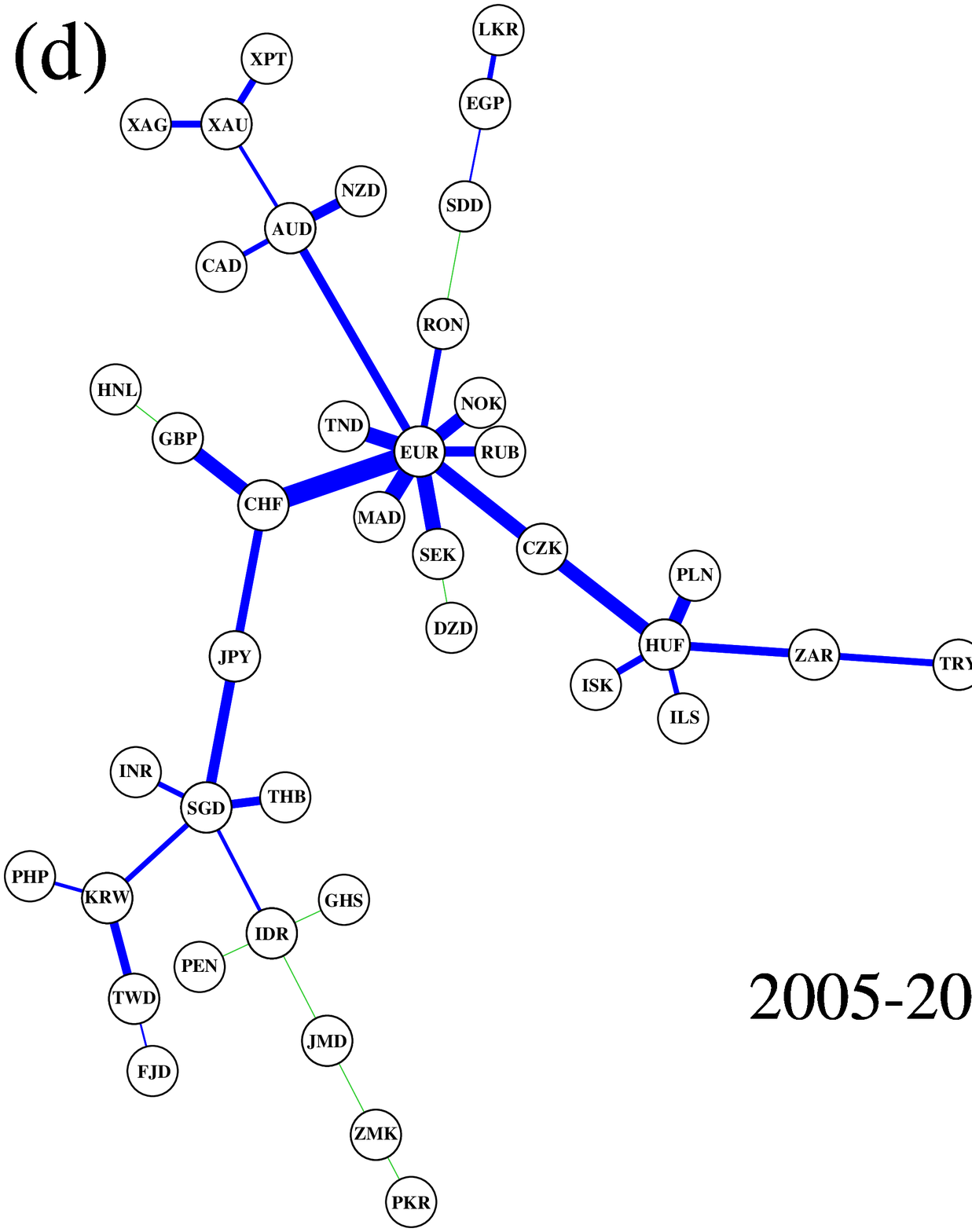}

\vspace{0.5cm}
\hspace{2.0cm}
\epsfxsize 5.5cm
\epsffile{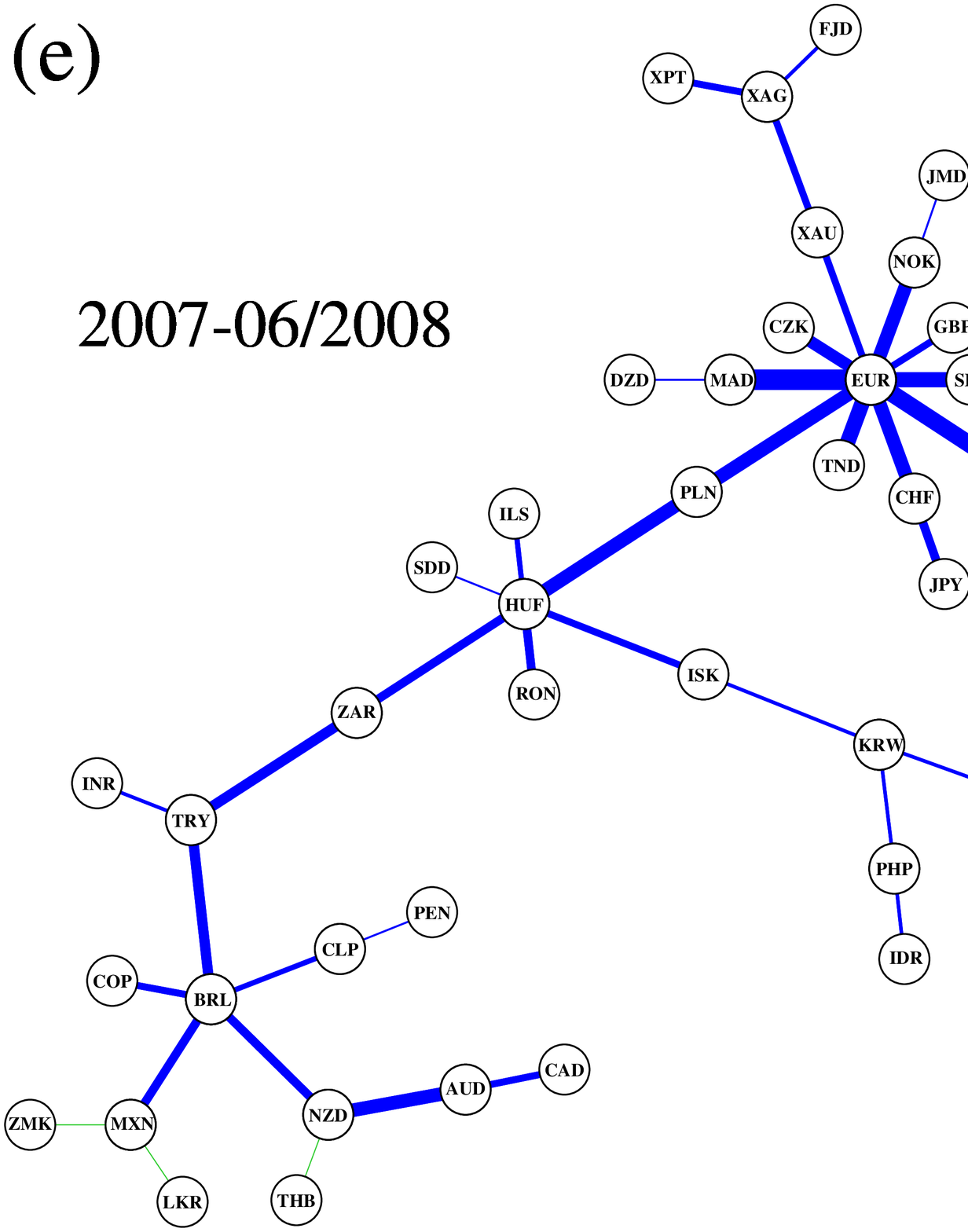}
\caption{Minimal Spanning Trees for the USD-based network calculated in   
subintervals of 2 years (a)-(d) and 1.5 year (e). Green edges connect 
anticorrelated nodes.}
\end{figure}

Another measure of how compact is the structure of a network is the 
characteristic path length quantifying the average minimal route between 
pairs of nodes. For an unweighted (binary) MST graph is given by a simple 
formula:
\begin{equation}
L^{\rm B} = {1 \over N (N - 1)} \sum_{\rm X,Y: X \neq Y} l^{\rm B}({\rm
X,Y}),
\label{pathlength}
\end{equation}
where $l^{\rm B}({\rm X,Y})$ denotes the number of edges in a path between
nodes X and Y. For the XAU-based MST: $L^{\rm XAU} = 1.65$.

The XAU-based network is a representation of the actual FX market as 
viewed from the perspective of a neutral observer. It therefore reflects 
only the primary, global structure of the market with two leading nodes:  
USD and EUR. In order to inspect finer details, we have to eliminate one 
of these currencies by choosing it as the base currency. The upper panel 
of Figure 2 shows MST for the EUR-based network and the lower one shows 
its counterpart for the USD-based network. These two trees have strikingly 
different topology. The EUR-based MST develops a single cluster of nodes 
with the USD node (of the degree $K=22$) in its centre. This USD-led 
cluster comprises currencies of Latin America, Asia, and the Mediterranean 
region.  Disintegration of the EUR-related cluster seen in Figure 1 caused 
a migration of such nodes as TND and MAD, normally evolving under the 
influence of euro, to their secondary attractor $-$ USD. Other nodes from 
that cluster are dispersed more or less randomly over the whole tree and 
they are connected via edges with small weights. With the topology that 
apparently resembles ($L^{\rm EUR} = 1.53$) the topology of the XAU-based 
network (Figure 1), the EUR-based network actually has on average evidently 
smaller edge weights and a smaller clustering coefficient ($\tilde{C}^{\rm 
EUR} = 0.33$).

\begin{figure}
\hspace{1.5cm}
\epsfxsize 10cm
\epsffile{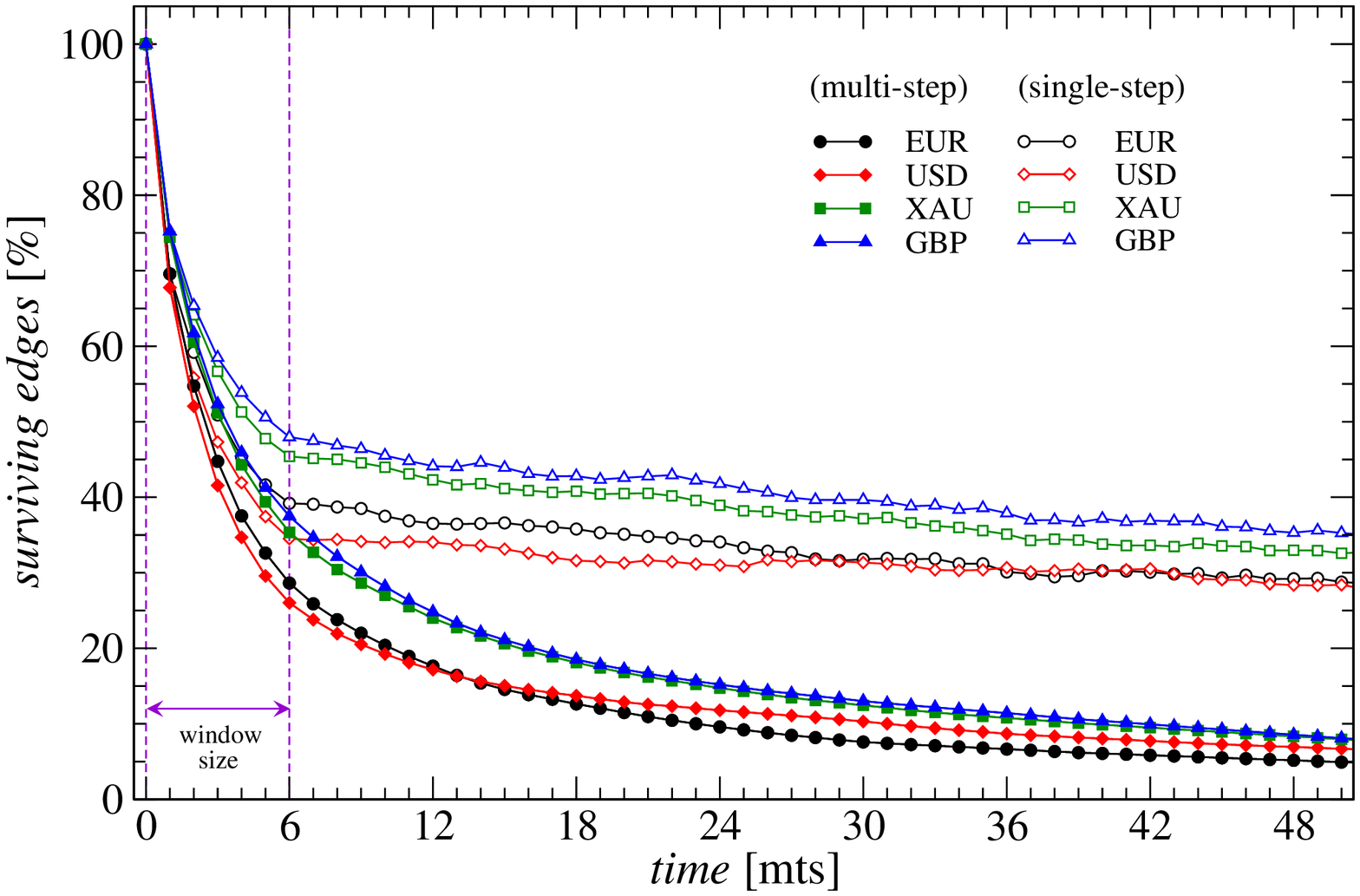}
\caption{Single-step survival ratio $\sigma^{\rm B}(\delta)$ (open 
symbols) and multi-step survival ratio $\Sigma^{\rm B}(\delta)$ (full 
symbols) for 4 representative base currencies: USD (red diamonds), EUR 
(black circles), GBP (blue triangles), and XAU (green squares). Width of 
the moving window is denoted by vertical dashed lines.}
\end{figure}

In contrast to both the XAU-based and the EUR-based minimal spanning 
trees, the USD-based MST (the lower panel of Figure 2) can be described as 
being somewhere between a hierarchical and a random graph. There is no 
central node, the node degree distribution does not have convincing 
scale-free tails (see ref.~\cite{gorski08} for more details), the 
characteristic path length is long ($L^{\rm USD} = 4.10$), and the clustering 
coefficient is small ($\tilde{C}^{\rm USD} = 0.11$). However, despite this 
small value of $\tilde{C}^{\rm USD}$, a few clusters can be identified in the 
MST: the European cluster concentrated around EUR, the South and East Asian 
cluster with SGD as its main hub, the commodity-trade-related cluster around 
AUD, and the Latin American cluster with the distinguished CLP node. A 
characteristic feature of this network is the existence of a number of nodes 
with rather insignificant couplings to the rest of the tree (the thinnest 
edges in the lower panel of Figure 2). That these edges are distributed almost 
randomly is suggested by a lack of causal relations between the underlying 
exchange rates and a lack of considerable economic ties between the 
corresponding countries. It is noteworthy that due to the US dollar's 
dominating role in the global financial system, the currency network based on 
USD has the finest possible cluster structure of all the network 
representations of the FX market.

Table 1 summarizes this part of our work, by collecting values of 
$L^{\rm B}$ and $\tilde{C}^{\rm B}$ for a few exemplary choices 
of the base currency. All the remaining network representations which are 
not shown here can be described by the values not exceeding the shown 
extremes for GHS and USD.

\begin{table*}
\begin{center}
\begin{tabular}[c]{|p{0.6cm}||p{0.8cm}|p{0.8cm}|p{0.8cm}|p{0.8cm}|p{0.8cm}|p{0.8cm}|p{0.8cm}|p{0.8cm}|p{0.8cm}|p{0.8cm}|}
\hline
& base: USD & base: GBP & base: EUR & base: MXN & base: CHF & 
base: AUD & base: PLN & base: JPY & base: XAU & base: GHS \\
\hline\hline
$L^{\rm B}$ & 4.10 & 2.33 & 1.53 & 2.14 & 1.63 & 2.44 & 1.99 & 1.95 & 1.65 
& 1.55 \\
\hline
$\tilde{C}^{\rm B}$ & 0.11 & 0.31 & 0.33 & 0.42 & 0.43 & 0.46 & 
0.51 & 0.51 & 0.71 & 0.93 \\
\hline
\end{tabular}
\caption{Characteristic path length $L^{\rm B}$ and average weighted clustering coefficient $\tilde{C}^{\rm B}$ for a few representative choices of the base currency B calculated for the full period 12/15/1998 $-$ 06/30/2008.}
\end{center}
\end{table*}

The networks discussed so far and presented graphically in Figures 1 and 2 
are in fact only the time-averaged representations of the real currency 
networks with constantly evolving structure. This evolution can be 
observed and quantified after increasing temporal resolution of our 
analysis. Let us split our time series of daily returns into the following 
subintervals: 1999-2000, 2001-2002, 2003-2004, 2005-2006, and 
2007-06/30/2008. For a given base currency we calculate the correlation 
matrix and the MST graph in each of the subintervals. Figures 3 and 4 show 
such MSTs for the EUR-based network and the USD-based network, respectively.

\begin{table*}
\begin{center}
\begin{tabular}[c]{|p{1.7cm}||p{0.8cm}|p{0.8cm}|p{0.8cm}|p{0.8cm}||p{0.8cm}|p{0.8cm}|p{0.8cm}|p{0.8cm}|}
\hline
& \multicolumn{4}{|c||}{ $\tilde{C}^{\rm B}$ } & \multicolumn{4}{|c|}{ $L^{\rm B}$ } \\
\cline{2-9}
& base: USD & base: GBP & base: EUR & base: XAU & base: USD & base: GBP & 
base: EUR & base: XAU \\
\hline\hline
1999-2000 & 0.08 & 0.35 & 0.43 & 0.72 & 4.57 & 1.70 & 1.21 & 1.25 \\
\hline
2001-2002 & 0.09 & 0.29 & 0.40 & 0.65 & 4.53 & 2.01 & 1.60 & 2.30 \\
\hline
2003-2004 & 0.15 & 0.34 & 0.32 & 0.66 & 3.34 & 2.62 & 2.51 & 2.47 \\
\hline
2005-2006 & 0.19 & 0.33 & 0.30 & 0.83 & 3.69 & 2.58 & 1.99 & 2.94 \\
\hline
2007-2008 & 0.19 & 0.37 & 0.27 & 0.80 & 3.68 & 3.67 & 2.92 & 2.67 \\
\hline
\end{tabular}
\caption{The weighted clustering coefficient $\tilde{C}^{\rm B}$ for the USD-, 
GBP-, EUR-, and XAU-based networks together with the characteristic path 
length $L^{\rm B}$ for the corresponding MSTs, calculated in subintervals of 2 
years (except the last one which is only 1.5 years long). Note the monotonic 
trends in $\tilde{C}^{\rm USD}$ and $\tilde{C}^{\rm EUR}$.}
\end{center}
\end{table*}

The trees in Figures 3 and 4 confirm the significant unstability of the
networks, reported already in ref.~\cite{mcdonald05,naylor07}. However,
the observed changes of the structure reveal also some slowly varying
components that survive for a few consecutive time intervals. Such a
component that can be found in the EUR-based MSTs is a decrease of the USD
node degree from $K=24$ in 1999-2000, through $K=18$ in 2001-2002, to
$K=13$ over the last 1.5 years (Figure 3). Also in terms of the node
betweenness $b^{\rm EUR}$, the USD node gradually loses its centrality:
0.88 (1999-2000), 0.86 (2001-2002), 0.83 (2003-2004), 0.82 (2005-2006),
and 0.71 (2007-2008). On the other hand, for the USD-based network the EUR
node does not change its degree so dramatically: the degree increased from
$K=6$ in 1999-2000 to $K=9$ in 2001-2002 and then it stabilized itself
around this value.

\begin{figure}
\hspace{1.5cm}
\epsfxsize 10cm
\epsffile{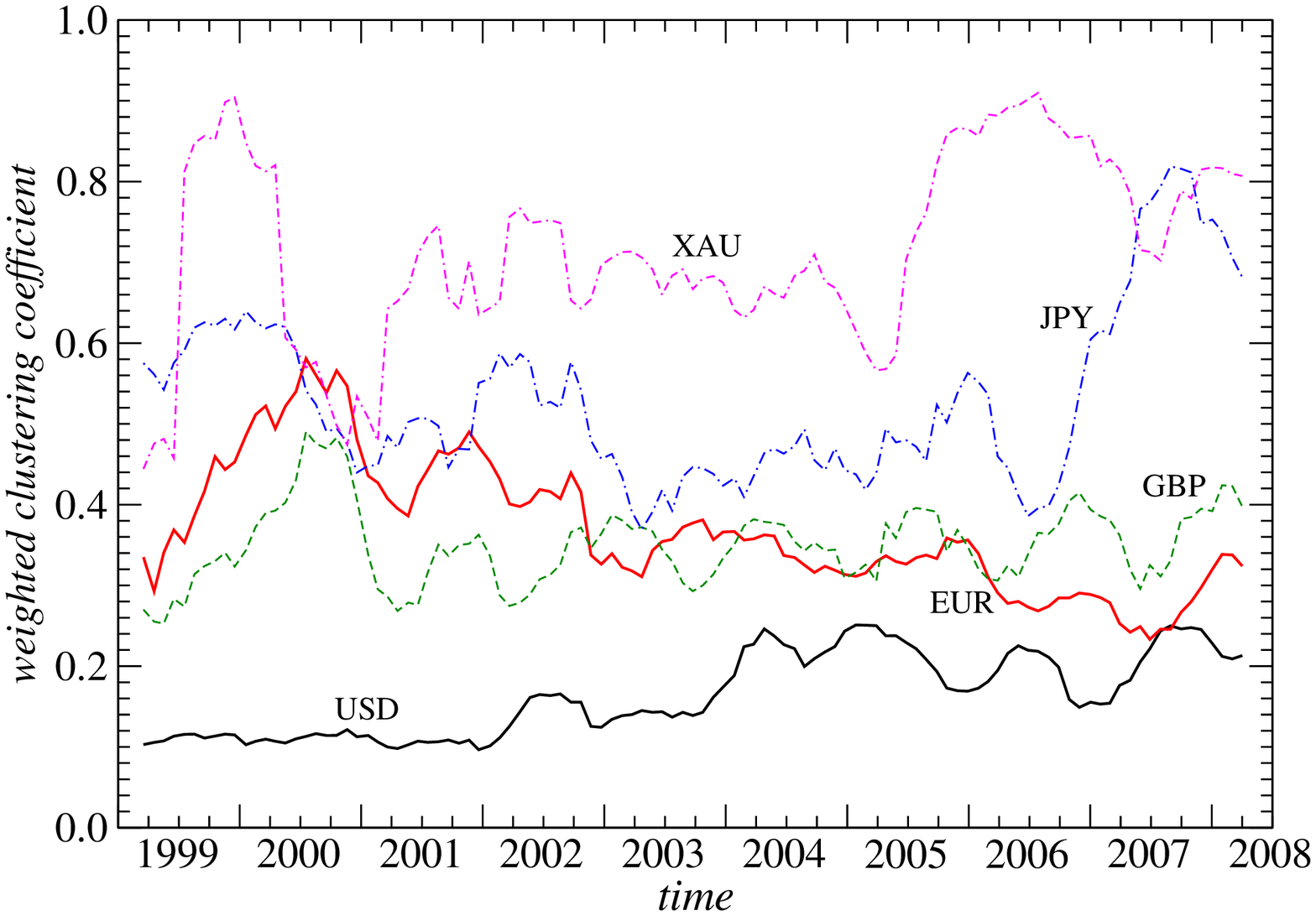}
\caption{Weighted clustering coefficient $\tilde{C}^{\rm B}$ as a 
function of time for 5 base currencies: USD (black solid), EUR (red 
solid), GBP (green dashed), JPY (blue dot-dashed), and XAU (magenta 
dot-double-dashed).}
\end{figure}  

In order to inspect how these changes affect the average topological
properties of the networks, we calculated values of the clustering
coefficient $\tilde{C}^{\rm B}$ and of the characteristic path length
$L^{\rm B}$ for four base currencies: USD, GBP, EUR and XAU. The 
results are collected in Table 2. Indeed, for EUR as the base currency the 
network has gradually become less clustered than it used to be in 
1999-2000 ($\tilde{C}^{\rm EUR}$ has dropped from 0.43 to 0.27), while the 
opposite can be said about the USD-based network ($\tilde{C}^{\rm USD}$ 
has increased from 0.08 to 0.19). For other choices of B situation is less
clear as we cannot identify such explicit trends. Furthermore, we observe
that the EUR-based, the GBP-based and the XAU-based MSTs are now more
dispersed than they used to be before (a strong increase of $L^{\rm B}$).
In contrast, the USD-based tree has become more compact (a noticeable
decline of $L^{\rm USD}$ in Table 2).

The existence of monotonic trends in the evolution of the network topology 
motivated us to track this phenomenon with a higher temporal resolution. 
We sampled the data with a moving window of length of 6 months (126 
trading days) and with a step of one month. As before, for each window 
position we calculated the corresponding correlation matrices and MSTs. 
For such short time intervals we expect that the inter-window fluctuations 
of the network structure are considerably stronger than in the previous 
case. Therefore we begin with quantitative estimation of the MST temporal 
stability by means of the single-step survival ratio
\begin{equation}
\sigma^{\rm B} (\delta) = {\# \{E^{\rm B}(i) \cap E^{\rm B}(i+\delta)\} 
\over N - 1},
\end{equation}
and the multi-step survival ratio~\cite{onnela05}
\begin{equation}
\Sigma^{\rm B} (\delta) = {\# \{E^{\rm B}(i) \cap E^{\rm B}(i + 1) \cap 
\dots \cap E^{\rm B}(i+\delta)\}
\over N - 1},
\end{equation}
where $E^{\rm B}(i)$ denotes a set of the B-based MST edges for a window 
$i$. These two ratios tend to overestimate and underestimate, 
respectively, the number of stable edges in the network~\cite{mcdonald05}. 
Figure 5 displays the average $\sigma^{\rm B} (\delta)$ and $\Sigma^{\rm 
B} (\delta)$ expressed in per cent units for different values of time 
shift $\delta$. It occurs that for disjoint windows ($\delta \ge 6$) the 
single-step survival ratio is a slowly decaying quantity indicating that, 
on average, roughly 1/3 of the edges in the initial network exist also 
after 4 years of evolution. In contrast, only about 5-10\% of the edges 
(i.e. 2-4 edges)  remain actually unchanged at least for 4 years. We see 
that the differences between the networks representing distinct base 
currencies are small. From Figure 5 one can infer that a vast majority of 
the edges change their locations with high frequency - as much as 60-75\% 
of the total number of edges, depending on B, do not survive for more than 
5 months. These numbers can be compared with the results reported 
in~\cite{mcdonald05}, where the analyzed network of 110 exchange rates 
between 11 currencies was considerably more stable and about 50\% of links 
were surviving for 2 years. However, the authors of ref.~\cite{mcdonald05} 
considered only the most liquid major currencies whose mutual relations 
are more stable than the relations involving less liquid currencies. 
Moreover, they analyzed data from a distinct time interval characterized 
by a lack of strong movements of the major cross-rates.

\begin{figure}
\hspace{1.5cm}
\epsfxsize 10cm
\epsffile{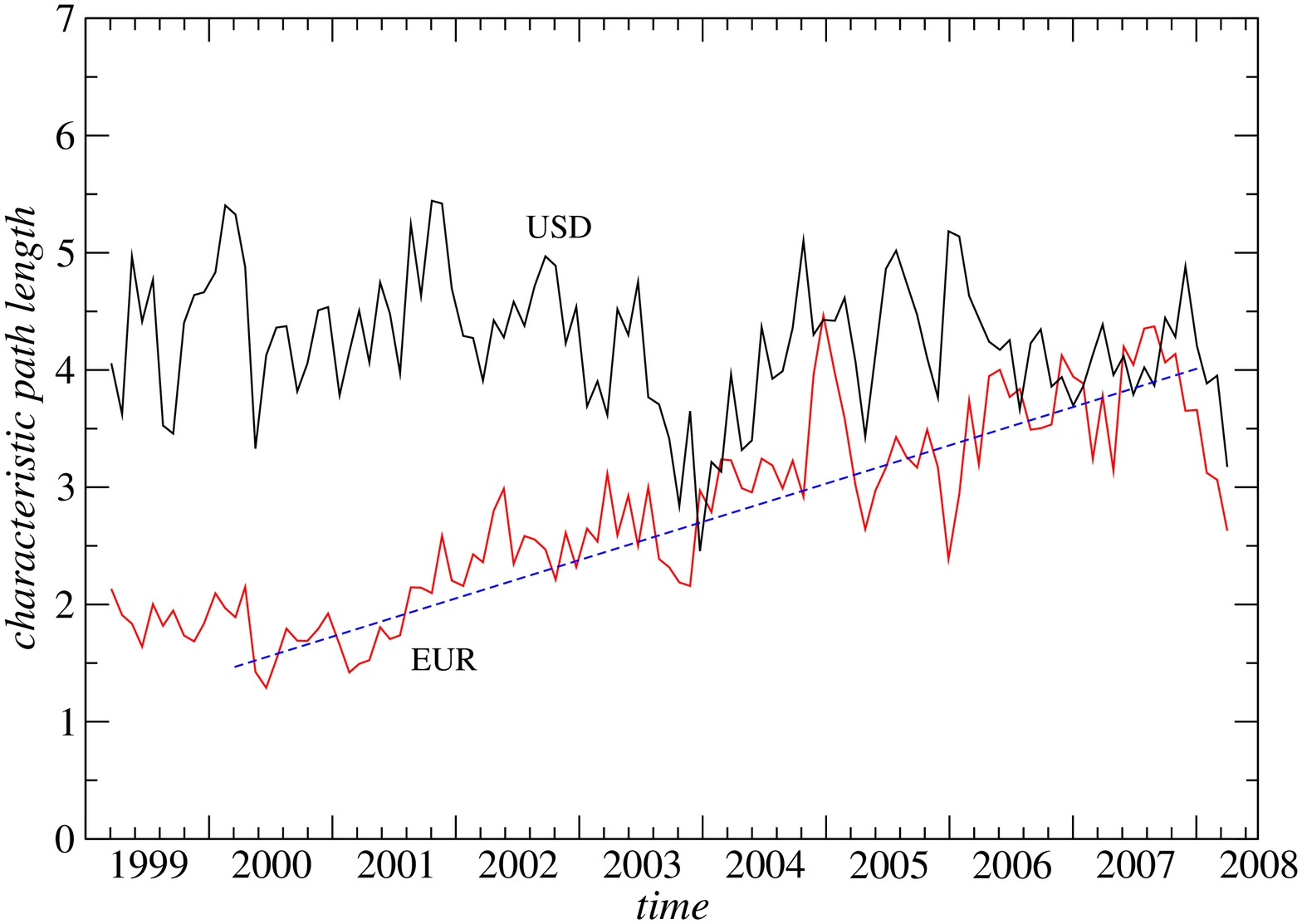}
\caption{Time dependence of characteristic path length $L^{\rm B}$ for the 
USD-based (black line) and the EUR-based MST (red line). Dashed line denotes linear trend in $L^{\rm EUR}$.}
\end{figure}

\begin{figure}
\hspace{1.5cm}
\epsfxsize 10cm
\epsffile{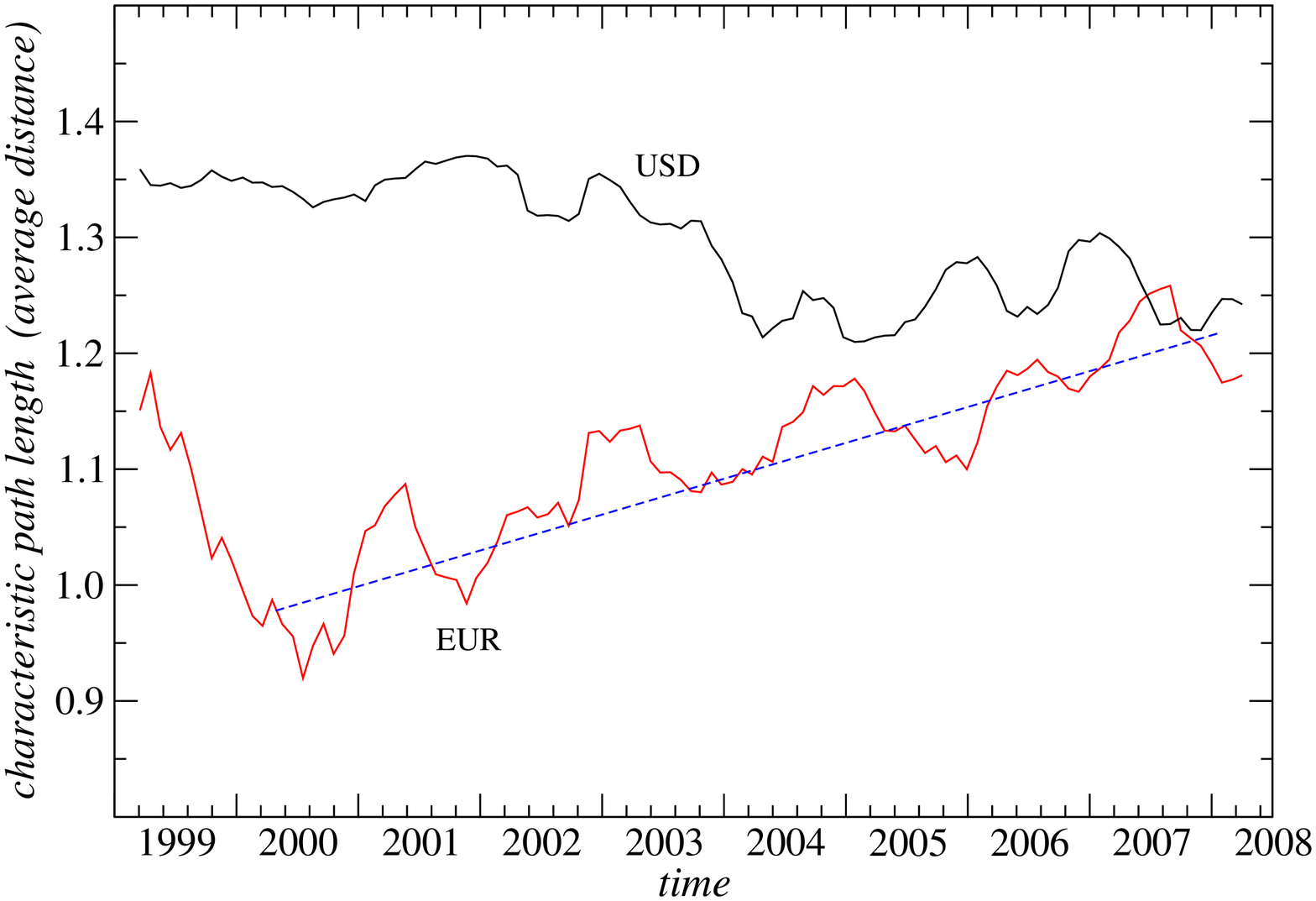}
\caption{Time dependence of the average internode distance 
$\mathcal{L}^{\rm B}$ for the USD-based (black line) and the EUR-based 
fully connected network (red line). Dashed line denotes linear trend 
in $\mathcal{L}^{\rm EUR}$.}
\end{figure}

Temporal evolution of the average weighted clustering coefficient for the
networks corresponding to USD, EUR, GBP, JPY, and XAU can be seen in
Figure 6. The most interesting behaviour is presented by the EUR-based
network.  After the introduction of euro at the end of 1998, during next
1.5 years $\tilde{C}^{\rm EUR}$ doubled its value from 0.3 to almost 0.6.
Then a long process of declining started, which drove the coefficient to
its absolute minimum below 0.25 around the middle of 2007. In the first
half of 2008 $\tilde{C}^{\rm EUR} = 0.33$, i.e. it returned to its initial
value. Since the short-time fluctuations of $\tilde{C}^{\rm EUR}$ occur
frequently, it is impossible to state whether at present we are witnessing
a trend reversal or the latest coefficient increase is only transient,
while we are still in the downward trend of $\tilde{C}^{\rm EUR}$.  
Nevertheless, over past 8 years the currency network viewed from the EUR
perspective evolved from a very centralized, USD-oriented network towards
a more dispersed, less clustered structure. On the other hand,
$\tilde{C}^{\rm USD}$ shows a more balanced behaviour. It started with a
value of about 0.1 which was considerably stable for 3 years. Then a
period of significant increase of $\tilde{C}^{\rm USD}$ up to 0.2 at the
end of 2003 followed, after which no further monotonic trends can be
identified. Since the beginning of 2004, $\tilde{C}^{\rm USD}$ has been
fluctuating around its average value of 0.2. This means that the USD-based
representation of the FX network, in which the EUR node plays a role of
the largest hub, has not changed its average structure significantly over
past a few years.

\begin{figure}
\hspace{1.5cm}
\epsfxsize 10cm
\epsffile{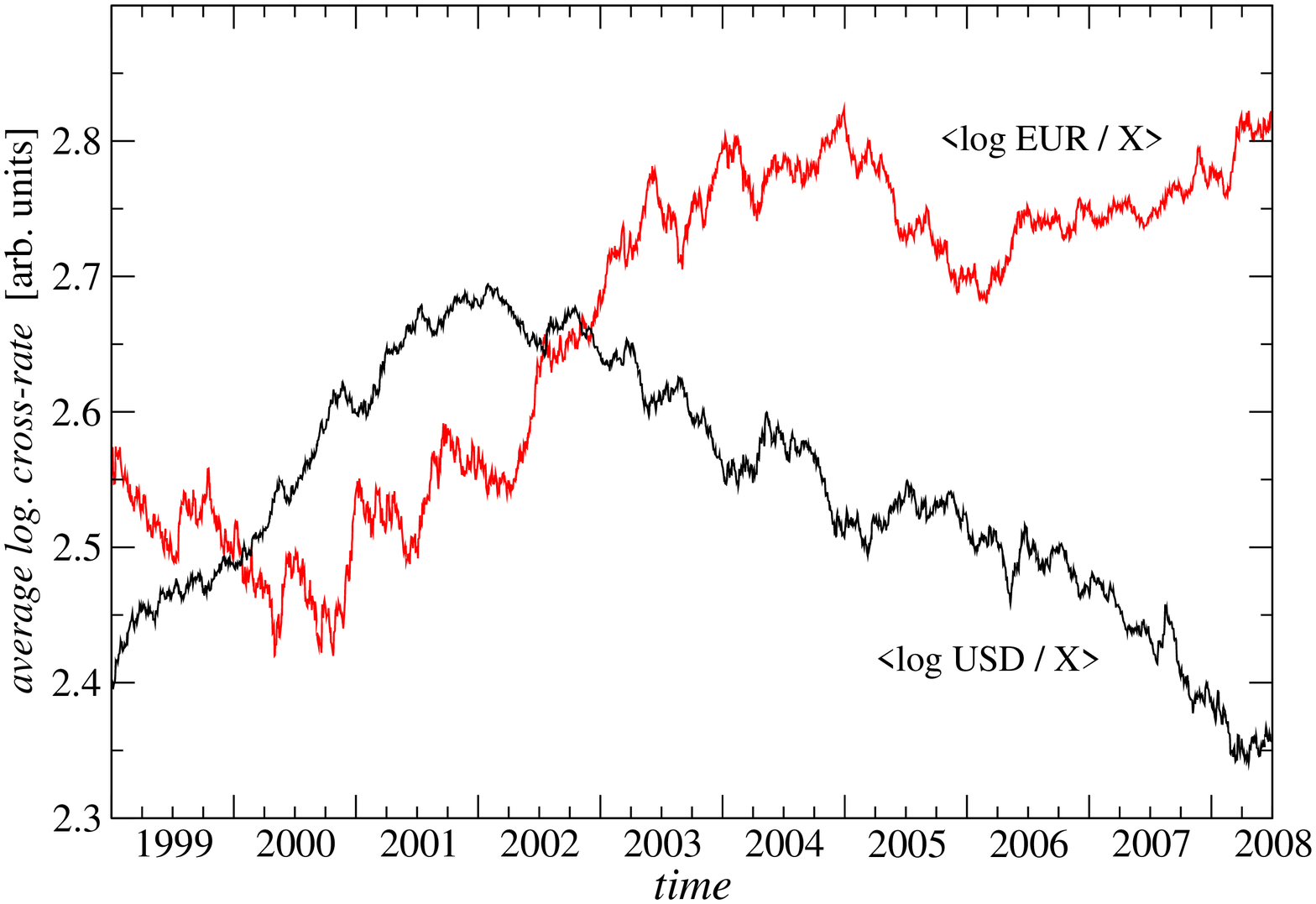}
\caption{Time dependence of logarithms of USD/X (black) and EUR/X (red) 
cross-rates averaged over all 45 price currencies X.}
\end{figure}

Out of three other base currencies from Figure 6, the GBP-based network
shows, on average, the highest stability, while both the JPY-based and the
XAU-based networks have largely unstable structure. It is interesting that
recently the network observed from the JPY perspective has similar
clustering coefficient as its XAU-based counterpart. It suggests that the
Japanese currency is now strongly decoupled from the rest of currencies
and has its own unique evolution. This coincides with a declining role of 
the Japanese currency in the international currency trade~\cite{bis07}.

The above conclusions for USD and EUR receive additional support from
Figure 7, where an evolution of the characteristic path lengths is
presented for two network representations. As it might be expected based
on the outcomes for the clustering coefficient, $L^{\rm EUR}(t)$ displays
a clear upward trend (approximated by a dashed line in Figure 7) which has
elevated $L^{\rm EUR}$ from about 1.5 in 2000 to about 4 in 2007. Also 
in agreement with the outcomes reported above, $L^{\rm USD}$ shows only 
rapid fluctuations around its long-time average, while we do not observe 
any trend.

These conclusions drawn from Figure 7, based on topological characteristics of 
the MSTs, can be confirmed by an analogous analysis carried out for the 
complete networks instead of their MST representations. In order to show this, 
we have to replace the definition~(\ref{pathlength}) of the characteristic 
path length $L^{\rm B}$ which, in the case of a fully connected network, would 
lead to a trivial result. We thus here adopt a definition based on weights 
rather than topology. According to this, the distance between two nodes is 
related to their coupling strength: the higher it is, the closer are the 
nodes. For the network analysed in our work, a good candidate for this measure 
is the metric distance $d^{\rm B}_{\rm X,Y}$ defined by Eq.(\ref{distance}). 
Then the average internode distance can be defined by
\begin{equation}
\mathcal{L}^{\rm B} = {1 \over N(N-1)} \sum_{\rm X,Y: x \neq Y} d^{\rm 
B}_{\rm X,Y}.
\end{equation}
Similar to $d^{\rm B}_{\rm X,Y}$, the above quantity can assume values in 
the range $0 \le \mathcal{L}^{\rm B} \le 2$ with a special case of 
$\mathcal{L}^{\rm B} = \sqrt{2}$ for a set of completely independent 
exchange rates. In Figure 8 we present temporal evolution of the average
internode distance $\mathcal{L}^{\rm B}$ for the USD-based and EUR-based
complete networks. We see that over the years $\mathcal{L}^{\rm EUR}$ has 
evolved according to an increasing linear trend which moved its value from
about 0.95 in 2000 to about 1.2 in 2007, in full analogy to what we observe
for MSTs in Figure 7. This similarity of behaviour of $L^{\rm B}$ and $
\mathcal{L}^{\rm B}$ can be considered an indication that minimal spanning
trees constitute a sufficient representation of the forex (or other financial) 
networks even though this type of graphs might seem to be rather arbitrarily 
chosen out of a variety of possible other choices.

\begin{figure}
\hspace{1.5cm}
\epsfxsize 10cm
\epsffile{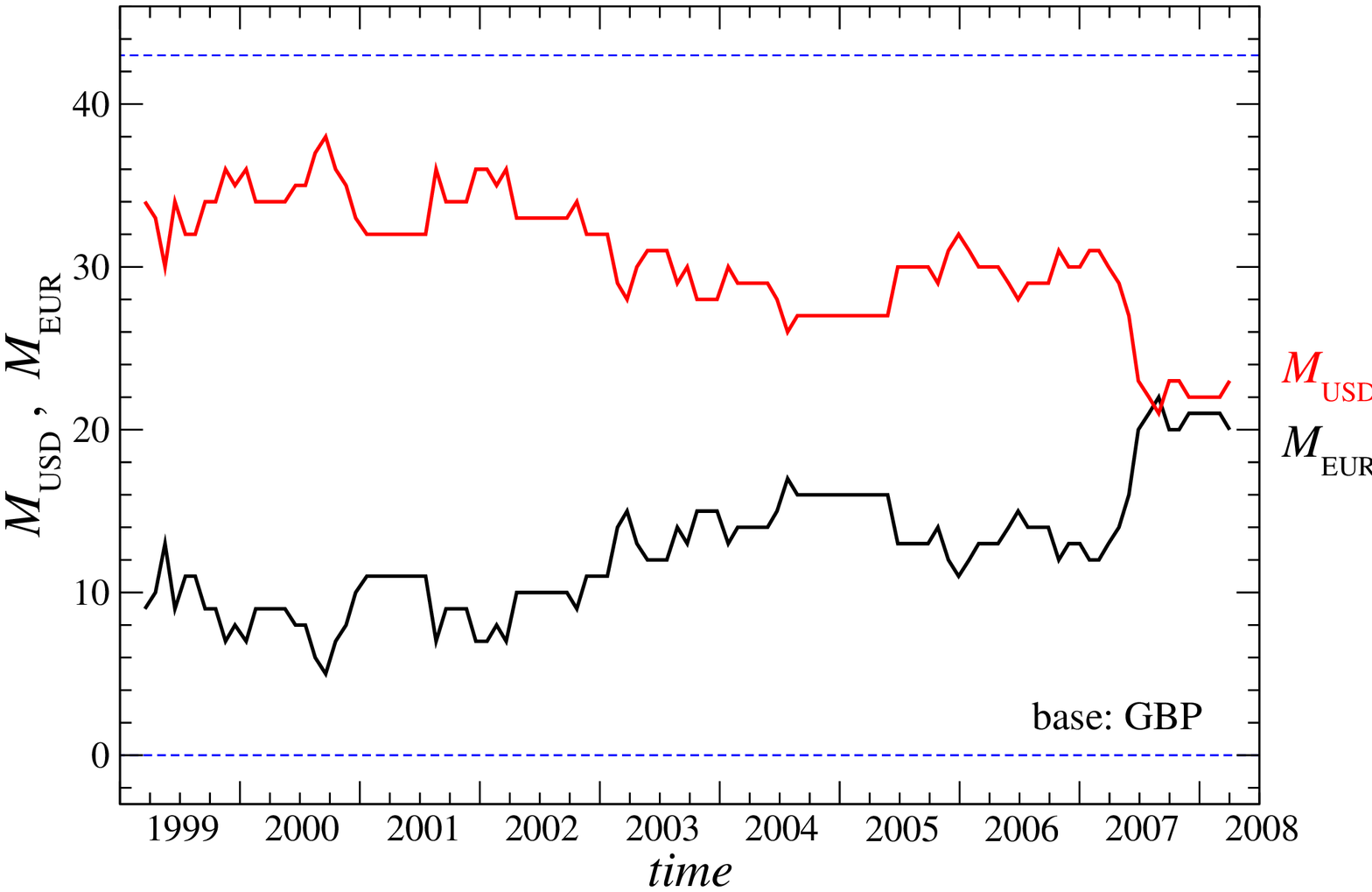}

\vspace{-0.5cm}
\hspace{1.5cm}
\epsfxsize 10cm
\epsffile{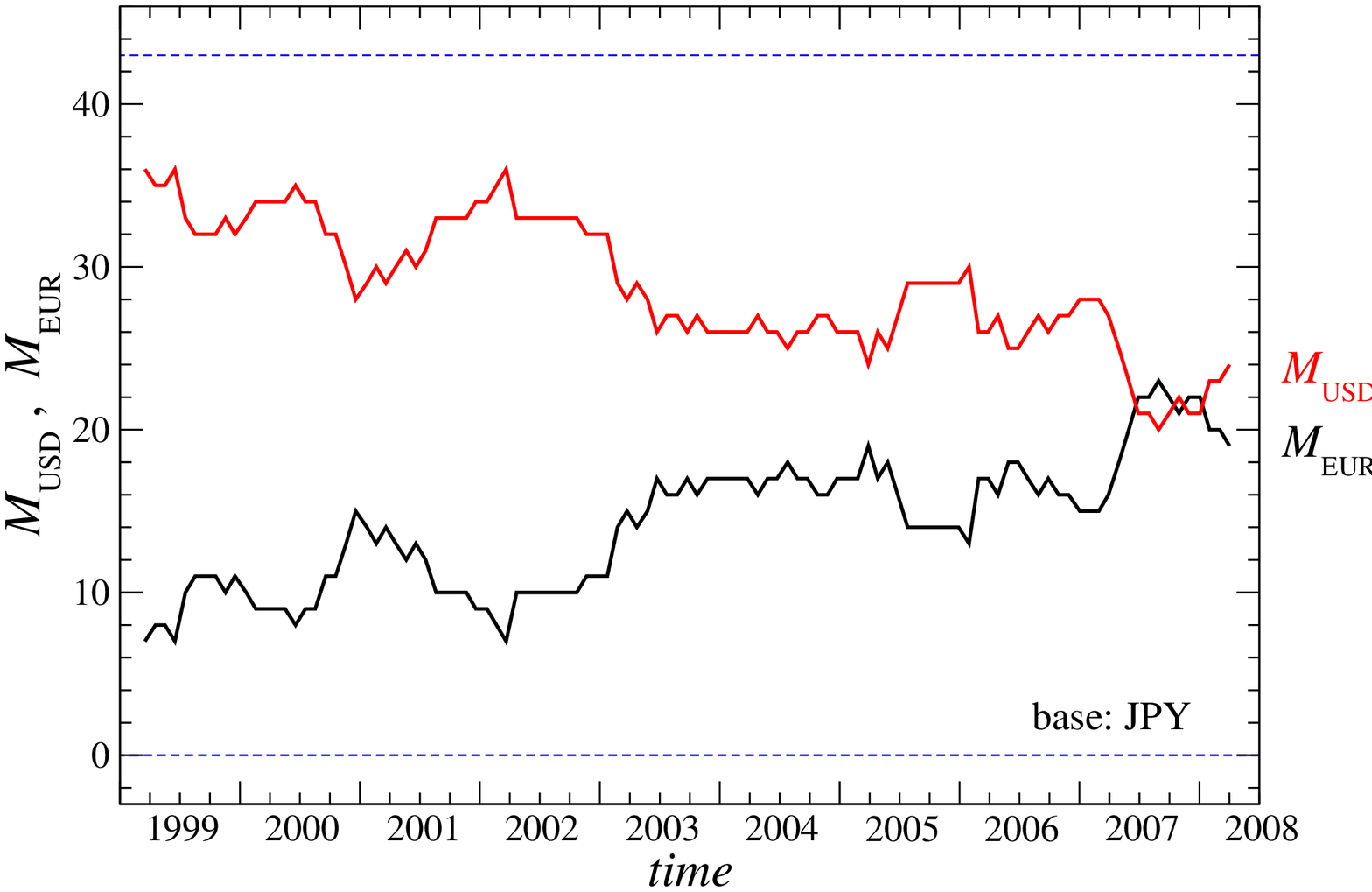}
\caption{Number $M_{\rm EUR}$ (black) of currencies closer to EUR than to 
USD in terms of the distance $d^{\rm B}_{\rm X,Y}$ (Eq.~(2)) and its 
complement $M_{\rm USD}$ (red) as functions of time for two representative 
choices of the base currency: GBP (top) and JPY (bottom).}
\end{figure}

Now a question emerges, what is the origin of the above-discussed evolution of 
the FX market. The decrease of the USD node's centrality, as expressed by a 
few different quantities, which is visible especially in the EUR-based network 
since 2000, reflects the fact that now fewer EUR-based cross-rates have a 
behaviour similar to the behaviour of the EUR/USD cross-rate. This 
phenomenon can have the following interrelated sources: (i) Some 
currencies, which originally were satellites of the US dollar (due to, 
e.g., strong economic dependence of the corresponding countries on the 
United States), could significantly release their ties and start a more 
independent evolution. Some of them might even become closer to EUR than 
to USD, what could explain the increase of the degree of the EUR node seen 
in Figure 4. (ii) Over the last years a strong depreciation of USD with 
respect to EUR and other currencies was observed: the EUR/USD cross-rate 
raised from 0.84 in July 2001 to 1.60 in April 2008. In this context it is 
worthwhile to look at the average cross-rates of USD and EUR with respect 
to other currencies shown in Figure 9. Since such a movement did not occur 
on all EUR-based cross-rates, its occurrence has weakened correlations 
between EUR/USD and some other EUR-based cross-rates. Thus, in the 
EUR-based representation of the FX network the USD node can now be less 
coupled to the rest of the network. This decrease of the USD couplings can 
also be observed from the perspectives of other currencies as Figure 10 
documents. It shows the number of currencies X whose GBP/X (JPY/X) 
cross-rates were more strongly correlated with GBP/USD (JPY/USD) than with 
GBP/EUR (JPY/EUR), denoted by $M_{\rm USD}$ (and its complement denoted by 
$M_{\rm EUR}$). Indeed, $M_{\rm USD}$ considerably declined between 2001 
and 2007. Similar observation can be made for a majority of alternative 
choices of the base currency.

\section{Summary}

We presented outcomes of our study of the currency network based on daily 
exchange rates of 46 currencies in the interval from the end of 1998 to 
the middle of 2008. We showed that the structure of the FX network depends 
on a choice of the base currency. In the case of currencies which are 
characterized by an independent, unique dynamics, as precious metals and 
some exotic currencies, the corresponding network representations have 
high node-node couplings and are described by high values of the average 
weighted clustering coefficient. A characteristic property is also almost 
complete lack of edges with small weights. These networks, if presented in a 
form of MST, have one dominating node of USD with a high node degree, which is 
a center of the largest cluster, and a few secondary nodes of a smaller 
degree. Apart from the USD-led cluster there exists also a significant cluster 
of European currencies. The USD node's dominance over the network is even more 
prominent in the network constructed for EUR and for a few other related 
currencies. For these currencies, however, due to the disintegration of the 
European cluster, the corresponding MST has a larger number of edges with 
small weights, attaching the associated nodes to sometimes random positions.
The clustering coefficient assumes medium values in this case. The third type 
of the FX network representations, with the USD-based network being its 
representative example, is described by the absence of a dominating 
node and by high values of the characteristic path length. The EUR node 
plays here a role of the most notable hub. Among other characteristic 
features of this representation there are: a clear geographically 
determined cluster structure, a large number of edges with small weights, 
and rather small values of the clustering coefficient. All other currency 
network representations can be located somewhere between the above three 
poles. It should be recalled, however, that despite these differences 
between the networks based on different currencies, all of them except the 
USD-based one have similar scale-free topology in terms of the node degree 
distribution~\cite{gorski08}.

The FX network is not stable in time and we showed that its evolution 
consists of at least two components. The first component is responsible 
for the rapid and unpredictable fluctuations of the network structure. It 
comprises both (i) the transient alteration of the cluster structure in 
which clusters are destroyed while new short-living ones emerge, 
indicating which currency is in play at the moment, and (ii) the wandering 
of individual nodes, visible in MSTs, due to random changes in correlation 
strengths between these nodes and the rest of the network. On the other 
hand, the second component is represented by slow variations of the 
network structure and it is responsible for the existence of metastable 
clusters and long-term trends. This kind of evolution is observed also in 
ecological networks, where large clusters, even if they build up and exist 
for a long period of time, in fact strongly fluctuate in size, leaving 
much freedom for noise (see, e.g., ref.~\cite{anderson05}). It is an open 
question if ecologically-motivated models can well describe properties of 
the currency market, since there are significant differences (like 
mutations) and between dynamics of species and currencies. Identifying 
similarities emerges as an intriguing issue for further study, however.

As our results show, there is a trend in the FX network's evolution, 
visible in all the network representations, which was determining the 
non-random modifications of the network structure for 7 years. It led to a 
significant decline of the node centrality of USD, as observed from the 
perpsective of a majority of the base currencies. We identified a possible 
source of this phenomenon in the strong depreciation of USD value with 
respect to other leading currencies. There is an additional - in fact 
related - possibility that the decrease of the USD node's importance with 
time and the increasing role of the EUR node, might be due to the fact 
that euro has become a more influential currency than it used to be in its 
early years and now more countries consider it as a sufficiently reliable 
alternative to the US dollar. It will thus be interesting to observe 
future evolution of the FX network, especially after stabilization of the 
USD cross-rates to other major currencies.

We finish with a remark on applicability of the minimal spanning trees to 
analyzing financial data. Despite the fact that MST is only one of many 
possible graphical representations of the actual network, in our opinion its 
topology can successfully be studied to collect information on the properties 
of the complete network. It should be noted that MST, due to the fact that its 
construction is based on selecting the strongest couplings among network 
nodes, comprises the core information on the global structure of the network. 
Thus, the overall characteristics of the network can typicaly also be 
reproduced in MST. From this point of view, we expect that even if one chose 
some other type of graph not being a tree, the conclusions based on them would 
be qualitatively similar to those inferred from the minimal spanning trees. In 
this context the simplicity of MST makes the application of this graph 
preferred. Another argument in favour of using MSTs is that results based on 
them are in satisfactory agreement with knowledge collected from other sources 
and from intuition.

\end{document}